\documentclass[
  twocolumn,english,aps,pra,
  superscriptaddress,amsmath,amssymb,floatfix
]{revtex4-2}

\usepackage{amsthm}
\usepackage{amsfonts}
\usepackage{siunitx}
\usepackage{amsmath}
\usepackage{amssymb}
\usepackage{graphicx}
\usepackage{verbatim}
\usepackage[colorlinks]{hyperref}
\usepackage{tikz}
\usepackage{pgfplots}
\usepackage{adjustbox}
\usepackage{braket}
\usepackage{xcolor}
\usepackage{physics}
\usepackage{amssymb} 
\usepackage{graphicx}
\usepackage{dcolumn}
\usepackage{bm}
\usepackage{mathtools}
\usepackage{hyperref}
\usepackage{mathrsfs}
\usepackage{dashrule}
\usepackage{caption}
\usepackage{subcaption}
\usepackage[normalem]{ulem}
\usepackage{color}

\definecolor{linkcolor}{RGB}{0,83,166}
\hypersetup{
  colorlinks = true,
  allcolors = {linkcolor}
}

\begin{document}

\title{Depth One Quantum Alternating Operator Ansatz as an Approximate Gibbs Distribution Sampler}

\author{Elijah Pelofske}
\email[]{epelofske@lanl.gov}
\affiliation{Information Systems \& Modeling, Los Alamos National Laboratory}
\affiliation{Quantum \& Condensed Matter Physics, Los Alamos National Laboratory}
\affiliation{Center for Quantum Computing, Los Alamos National Laboratory}

\begin{abstract}
This study numerically investigates the thermal sampling properties of QAOA, the Quantum Alternating Operator Ansatz which was generalized from the Quantum Approximate Optimization Algorithm. Specifically, we examine the ability of QAOA to sample from the Gibbs distribution, equivalently the Boltzmann distribution, defined by a classical Ising model, on a fully connected disordered spin glass (Sherrington-Kirkpatrick) model. We focus on two different QAOA mixers: the standard transverse field X mixer, and the Grover mixer. At a QAOA depth of one we examine, for a single full QAOA parameter search space period, the energy landscape, the Shannon entropy landscape of the QAOA probability distribution, and the tradeoff between Boltzmann distribution sampling temperature and error rate (how close to the true Boltzmann distribution is the QAOA distribution). We find that at very high temperatures one-round Grover mixer QAOA can sample from the Boltzmann distribution more accurately than the standard X mixer QAOA at one round. Both X mixer and Grover mixer depth one QAOA can serve as approximate Boltzmann distribution samplers, and the accuracy depends heavily on the QAOA angle choice. Moreover, depth one QAOA parameters that minimize the cost Hamiltonian energy do not correspond to the most accurate (approximate) thermal state samplers.

\end{abstract}

\maketitle

\section{Introduction}
\label{section:introduction}

Accurate and efficient sampling from a thermal distribution (Gibbs sampling) is a foundational computational capability for information processing tasks ranging from combinatorial optimization, to statistical mechanics, to machine learning, to Bayesian statistics, and even to computational biology~\cite{goto2018boltzmann, ponty2008efficient, flajolet2007boltzmann, augustin1996autologistic, rohl2004protein}. Gibbs sampling is in general a computationally hard problem, and any algorithmic improvement for more efficiently or accurately performing Gibbs sampling is of interest. Gibbs sampling is fundamentally connected to computational sampling tasks such as solving NP-hard combinatorial optimization problems. In this study, we consider the task of evaluating how well the Quantum Approximate Optimization Algorithm~\cite{QAOA, farhi2015quantum} (QAOA) can sample from a Boltzmann distribution defined by a disordered classical Ising model. QAOA was later generalized to include more complicated quantum driving mechanisms, dubbed the Quantum Alternating Operator Ansatz~\cite{Hadfield_2019}. This includes other mixing, or state transition, operations such as the XY mixers~\cite{PhysRevA.101.012320} and the Grover mixer~\cite{Bartschi_2020}. Our aim is to quantify how well QAOA can sample from a particular type of complex probability distribution, in this case a Boltzmann distribution which is defined as

\begin{equation}
    p(z) \propto e^{-\beta E(z)}, 
\end{equation}

where $z$ is a vector of spins, $E(z)$ is the energy of that spin configuration evaluated on an Ising model, and $\beta$ is the inverse temperature $\beta = 1/k_{B}T$. We will use natural units of the Boltzmann constant $k_B=1$ for plotting and numerical simplicity. 

This general sampling task of thermal state sampling goes by different terms in different fields of study. Gibbs sampling, Boltzmann sampling, and thermal sampling, all refer to the same sampling task, at least in this specific case where we are sampling from a classical Hamiltonian. In the text we will typically use the term Gibbs sampling.

The key motivation for this study is twofold. First, quantum algorithms such as QAOA can be implemented as very short-depth quantum circuits, making them practical to implement, albeit with noise, on current quantum computers~\cite{He_2025, Harrigan_2021, Pelofske_2024_scaling_QAOA, pelofske2023short, shaydulin2023qaoancdotpgeq200, Pelofske_2023, Weidenfeller_2022}. Therefore, if QAOA can be a good sampler from complex probability distributions, such as Boltzmann distributions, even for moderate circuit depth (e.g., relatively small $p$), then those circuits could be run on current quantum computers. Second, different variants of QAOA have different thermodynamic sampling properties. Namely, Grover mixer QAOA~\cite{Bartschi_2020} samples spin configurations that have the same cost value entirely uniformly, whereas X mixer QAOA can sample configurations of the same cost non-uniformly~\cite{Golden_2022_fair, Pelofske_2021, Pelofske_2025}, which is a similar property to continuous time transverse field driven quantum annealing~\cite{Albash_2015, matsuda2009quantum, PhysRevA.100.030303, PhysRevLett.118.070502, PhysRevE.99.063314}. In principle, this means that GM-QAOA (Grover-Mixer QAOA)~\cite{Bartschi_2020} may be a better thermal sampler, at least for some types of more frustrated Ising models, compared to the standard X mixer QAOA. In summary, GM-QAOA guarantees uniform probabilities for configurations at the same energy level of a classical Hamiltonian. This is an entropic property which is required for good thermal sampling (which is not strictly unique to GM-QAOA, other QAOA variants can have this property under certain conditions), but is not sufficient by itself for good thermal state sampling or preparation. Therefore, it is reasonable to analyze the differences between these two QAOA variants.

Sampling applications are one of the domains where it is plausible that quantum computers will be able to significantly outperform classical algorithm counterparts -- in the case of QAOA, there is evidence that for some QAOA trajectories and some cost functions, the output distributions are hard to approximate classically~\cite{farhi2019quantumsupremacyquantumapproximate, krovi2022averagecasehardnessestimatingprobabilities}. An interesting open question is how the classical approximability of a QAOA probability distribution connects to how well that QAOA probability distribution approximates a Boltzmann distribution.

QAOA is defined by several algorithmic components, which we will briefly summarize here. First, a cost function $C(\vec{z})$ encodes an optimization problem we wish to solve, by finding a globally optimal assignment of spins $\vec{z} = (z_1, z_2, \dots, z_N)$, where $z\in\{+1,-1\}^n$. We will sample solutions to this optimization problem using QAOA, which involves the following components.

An initial state $\ket{\psi}$.
A phase separating Hamiltonian $H_P$, which provides a phase shift to each basis state $z$ according to their cost value given by $C(\vec{z})$. In other words, $H_P$ encodes $C(\vec{z})$.
A mixer Hamiltonian $H_M$ which provides parameterized interference between basis states $z$. $H_M$ is effectively the driving mechanism of QAOA by facilitating state transitions. 
An integer $p \geq 1$. This determines the total number of rounds of simulation that are performed. Note that this parameter goes by many different terms in the literature, including \emph{rounds}, \emph{layers}, \emph{levels}, and \emph{depth}. 
A vector of real-values $\vec{\beta} = (\beta_1,...,\beta_p)$, of length $p$, which parameterizes $H_M$.
A vector of real-values $\vec{\gamma} = (\gamma_1,...,\gamma_p)$, of length $p$, which parameterizes $H_P$.

QAOA then performs an alternating Hamiltonian simulation of these two non-commuting Hamiltonians $H_P$ and $H_M$, each layer of these Hamiltonians generating a parameterized quantum state, as 

\begin{equation} 
\label{equation:QAOA_evolution}
\ket{\vec{\beta},\vec{\gamma}} = e^{-i\beta_p H_M} e^{-i\gamma_p H_P} \cdots e^{-i\beta_1 H_M} e^{-i\gamma_1 H_P}\ket{\psi}. 
\end{equation}

The state $\ket{\vec{\beta},\vec{\gamma}}$ is then measured in the computational basis to obtain a single sample, which is a candidate solution of the specified combinatorial optimization problem. Combined, $\vec{\beta}, \vec{\gamma}$ are typically referred to as the QAOA \emph{angles}. 
There is an important connection between the continuous time adiabatic quantum computation~\cite{farhi2000quantumcomputationadiabaticevolution}, and discretized digital protocols such as QAOA~\cite{Sack_2021, boulebnane2025equivalencequantumapproximateoptimization, mbeng2019quantumannealingjourneydigitalization, Kovalsky_2023, Monta_ez_Barrera_2025} -- namely that fundamentally the principle being used is that long evolution times will result in high solution quality (in the case of optimization problems, this is finding globally optimal solutions, or at least close to the global minimum). In the case of QAOA, this means the number of layers $p$ must be large in order to approach the global minimum, proportional to the system size.

Here, we will use the standard initial state of $\ket{\psi} = \ket{+}^{\otimes n}$ where $\ket{+} = \frac{1}{\sqrt{2}}(\ket{0} + \ket{1})$, which provides a uniform superposition over all basis states. And we will use two different $H_M$ Hamiltonians. The first is the standard transverse field, or X, mixer~\cite{QAOA, farhi2015quantum}; $H_M = \sum_{i=1}^N X_i$. The other mixer we will use is the Grover mixer~\cite{Bartschi_2020}, which is based on the Grover search algorithm~\cite{grover1996fastquantummechanicalalgorithm}. The computation that we evaluate in this study is whether QAOA can sample from Gibbs distributions of the phase separating Hamiltonian $H_P$, in other words, whether a low-depth QAOA parameter regime exists that can approximate a Gibbs distribution of $H_P$ at any effective temperature.

Note that there is a symbolic re-use collision in the thermodynamic and QAOA notation; $\beta$ denotes both the QAOA mixer parameter set, and the inverse thermodynamic temperature. So as to avoid confusion, we will use $\vec{\beta}$ or $\beta_{\text{p=1}}$ to refer to the QAOA parameter, and we will use either $\beta$ or $\beta_{\text{eff}}$ to refer to the inverse temperature of a probability distribution.

\subsection{Literature review and prior studies}

There have been a series of studies recently which have demonstrated using analog quantum computers as physical noisy samplers of a Boltzmann distribution (both quantum and classical)~\cite{Ball_2023, PhysRevApplied.11.044083, Izquierdo_2021, Marshall_2017, nelson2021singlequbitfidelityassessmentquantum, PRXQuantum.3.020317, PhysRevApplied.17.044046, buffoni2020thermodynamics, mörstedt2024rapidondemandgenerationthermal, sandt2023efficient, Wild_2021}. In general, there is quantum algorithmic design interest in whether quantum computers can accelerate thermal sampling computational tasks~\cite{Holmes2022quantumalgorithms, Poulin_2009}. The aim of this study is to examine in detail the thermal sampling properties of a very short depth digital quantum algorithm; depth one QAOA. 

Refs.~\cite{D_ez_Valle_2023_PRL, D_ez_Valle_2024, diezvalle2025universalresourcesqaoaquantum} have established standard QAOA can produce ``pseudo-Boltzmann'' states, specifically ref.~\cite{D_ez_Valle_2023_PRL} used a slightly modified standard X mixer QAOA circuit to generate pseudo-Boltzmann distributions for several different types of Ising models using one QAOA round. Ref.~\cite{PhysRevA.108.042411} did a similar type of analysis, arguing that the standard X mixer QAOA produces approximate Boltzmann distributions -- which was in part argued in the form of similar Shannon entropy between the ideal Boltzmann distribution and QAOA probability distributions. Ref.~\cite{Leontica_2024} compared QAOA thermal sampling and instantaneous quantum polynomial circuit thermal sampling. In the present study, the aim is to build on this line of inquiry by additionally comparing Grover mixer QAOA probability distributions. One of the ways in which the current study performs this analysis is using a configuration (sample) level probability thermal state analysis, not just by energy level density (a type of density-of-states) analysis, which gives a highly comprehensive sampling characterization using numerical computations. Additionally, we aim to thoroughly probe the accuracy tradeoff of depth one QAOA distributions -- in other words, to quantify how close to true Boltzmann distributions, of any temperature, can these QAOA pseudo-Boltzmann states get. 

Ref.~\cite{abe2026fairsamplingtemperaturetargetedqaoa} proposed a QAOA inspired algorithm, based on Ref.~\cite{PhysRevLett.99.030603}, which is able to compute very low-error approximations of classical Gibbs distributions at selected temperatures, although requiring fairly high circuit depths to do so (up to $10^2$). 

Although not specifically QAOA focused, there have been several related studies on more general variational algorithms to approximate thermal distributions. Ref.~\cite{Wu_2019} proposed a variational quantum algorithm, motivated by QAOA, to prepare thermal states of different Hamiltonians including Ising models. Ref.~\cite{warren2022adaptivevariationalalgorithmsquantum} proposed a variational quantum algorithm that aims to approximate quantum Gibbs distributions. Ref.~\cite{nakano2025neuralnetworkassistedmontecarlosampling} used a neural network to produce approximate Boltzmann distributions, which had been trained on QAOA output distributions.

Note that this line of research is distinct from the recently proposed quantum circuit based sampler as a component of the standard iterative Markov chain Monte Carlo algorithm~\cite{metropolis1953equation} where the sampler that is used is a short depth Trotterized quantum circuit~\cite{Layden_2023, Nakano_2024}, equivalent to QAOA (e.g., similar idea to digitized quantum annealing~\cite{mbeng2019quantumannealingjourneydigitalization}). In the present study, instead we investigate how well QAOA, using two different types of mixing operations, works for direct estimation of a Boltzmann distribution defined by an Ising model. In other words, this work is a numerical simulation study of the thermal sampling properties of QAOA, using two different types of non-commuting quantum Hamiltonian drivers; the standard transverse field mixer, and the Grover mixer.

\section{Methods}
\label{section:methods}

The Ising model we focus on primarily in this study is a single instance of a $15$-spin Sherrington-Kirkpatrick~\cite{PhysRevLett.35.1792} (SK) model. For all edges in the fully connected $15$-node graph, we randomly select either $+1$ or $-1$ for each local field $h_i$ and either $+1$ or $-1$ for the $J$ coupling, resulting in the spin glass Ising model of

\begin{equation}
    \mathcal{H} = \sum_{i < j} J_{ij} \sigma_i^z \sigma_j^z + \sum_i h_i \sigma_i^z, 
    \label{equation:SK_model_ising}
\end{equation}

where $\sigma_i^z$ are spins of either $\uparrow$ or $\downarrow$, encoded as qubits in QAOA. Typically the goal of combinatorial optimization is to minimize the energy of this classical Hamiltonian; in other words, find the variable assignment of the $\sigma_i^z$ spins which results in a globally optimal minimum energy. The specific Ising model used in this study has a globally minimum energy of $-44$, and a maximum energy of $+44$. Only a single representative Ising model is used in order to focus on and rigorously evaluate the thermal sampling properties of the QAOA across a high-resolution QAOA angle parameter search space. Considering a random representative problem instance is a reasonable choice in particular for studying the QAOA search landscape and sampling properties, of the SK model, because it is well known that QAOA control parameters concentrate strongly and perform very similarly for different instances of problems from the same class, including similar problems of very different size~\cite{PhysRevA.104.L010401, brandao2018fixedcontrolparametersquantum, boulebnane2021predictingparametersquantumapproximate, Shaydulin_2023, katial2024instancedependenceoptimalparameters, Pelofske_2024_scaling_QAOA}. This phenomenon goes by many different terms, including parameter concentration, parameter transfer, fixed point QAOA, and fixed angles, and has been analyzed on many different types of problem instances including specifically for SK models~\cite{LIPICS.TQC.2022.7, sakai2024linearlysimplifiedqaoaparameters, Farhi_2022, galda2021transferabilityoptimalqaoaparameters, galda2023similaritybasedparametertransferabilityquantum, chernyavskiy2025improvingqaoaapproximatequbo, PhysRevA.104.052419}. Therefore, it is reasonable to study the sampling properties of QAOA on small problem instances and we expect that these properties will approximately hold, due to QAOA control parameter concentration, for larger problem sizes for problems of similar types (in this case, disordered Ising models, specifically the SK model). For large problem instances, and over large ensembles of problem instances, QAOA parameters concentrate, but locally and for single problem instances QAOA properties can differ from the ensemble behavior. Importantly, QAOA parameter landscapes concentrate, not just optimal QAOA angles~\cite{brandao2018fixedcontrolparametersquantum, rudolph2021orqvizvisualizinghighdimensionallandscapes, Farhi_2022, Kr_ger_2025, Claes_2021}. QAOA parameter transfer also applies to the distribution~\cite{Lykov_2023}, not just the average cost-function landscape, including specifically for the SK model~\cite{Farhi_2022}. In this study, although the focus is thermodynamic sampling, the relevant sampling is exactly the QAOA produced probability distribution in the computational basis. Therefore, good thermal sampling QAOA parameters follow the same concentration characteristics as any other QAOA parameter set. In order to include some additional problem instances at other sizes, Appendix~\ref{section:10_spin_results_appendix} and \ref{section:5_spin_results_appendix} report $10$ and $5$ spin SK model results as well, although in the main text we focus on one $15$-spin instance. The results on different problem sizes show that the QAOA parameter landscapes concentrate, for both X and Grover mixer. These SK Ising models are encoded as the phase separator Hamiltonian in the QAOA simulation.

The quantum simulation tool which we use to compute all probabilities and expectation values is the Julia~\cite{bezanson2017julia} based \texttt{JuliQAOA} simulator~\cite{Golden_2023}, which is specifically tailored for exact QAOA numerical simulation. The numerical simulations we consider are noiseless, and not represented in terms of compiled circuit models, but instead directly in the form of Hamiltonians. No qubit depolarizing error sources, or other types of error sources, are considered, and no shot noise (finite sampling effect) is considered either. The goal of this is to emphasize the noiseless thermal properties of the approximate quantum optimization algorithms. Considering finite sampling effects is an important line of study for future research.

The goal is to characterize whether certain probability distributions, in this case QAOA distributions which are parameterized by $\gamma_1$ and $\beta_1$, result in Boltzmann distributions -- or, approximations of Boltzmann distributions. However, we do not in general know what temperature QAOA may be a good Boltzmann sampler at. To this end, we perform a fitting process for different $\beta$ values in order to find which value results in the lowest error. This process is performed using a combination of a $\beta$ parameter gridsearch and black-box optimizers (e.g., no derivative or gradient is known a priori). This black-box optimization process still involves some statistical error, so accordingly this specifically means that there could be some $\beta$ which actually gives a lower error rate than what we quantify. However, the methods we employ here attempt to minimize this statistical error as much as is reasonably possible. This caveat relates primarily to the precision of the determined  $\beta$ which gives the lowest error rate; this optimization procedure is not guaranteed to deterministically find the optimal $\beta$, to arbitrary precision for each QAOA probability distribution. However, the search landscapes are not incredibly complex and hard to optimize over (Appendix~\ref{fig:example_TVD_beta_response_surfaces} reports several examples of these response surfaces), which means that in general we expect this optimization procedure to be reasonably accurate. The goal of this type of study is a numerical implementation study using classical simulation of digital quantum circuits. However, the corresponding implicit goal is to eventually run these types of algorithms on low-error actual quantum computers. To this end, on actual quantum computers issues such as shot noise will necessarily be part of the computation. Therefore, although this study does not examine finite sampling effects, there are additional real quantum computer considerations that would need to be addressed in future work.

We need an error rate to quantify how far the QAOA probabilities are from an ideal Boltzmann distribution. To this end, there are many different error measures that could be used. We will use total variation distance (TVD) which is a simple, intuitive, and bounded error measure defined for two probability distributions, $P(x)$ and $Q(x)$ as

\begin{equation}
\text{TVD} = 0.5 \sum_x |P(x) - Q(x)|. 
\label{equation:TVD}
\end{equation}

\begin{figure*}[ht!]
    \centering
    \includegraphics[width=0.495\linewidth]{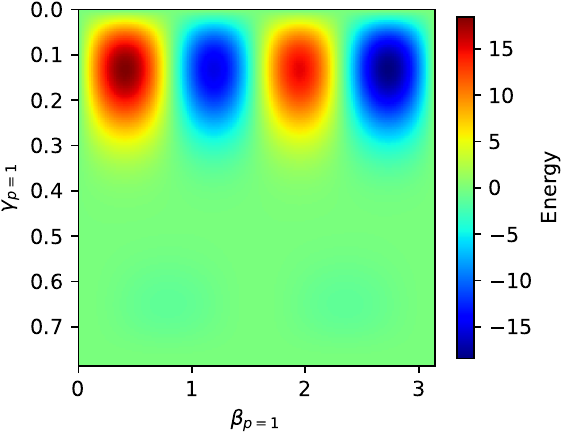}
    \includegraphics[width=0.495\linewidth]{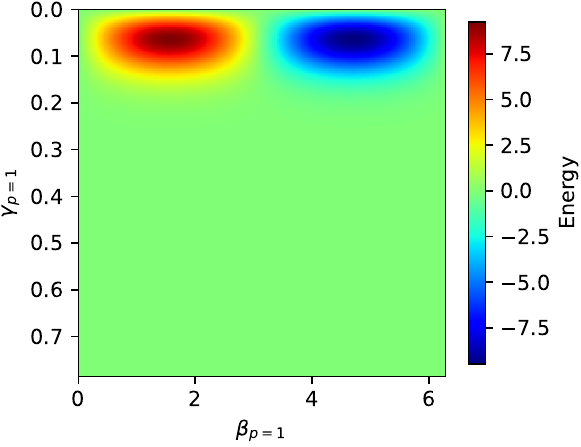}
    \caption{Energy $p=1$ QAOA angle landscape for the X mixer (left) and the Grover mixer (right). Recall that $\beta_{p=1}$ here is a QAOA angle parameter, not inverse temperature.  }
    \label{fig:QAOA_energy_heatmap}
\end{figure*}

\begin{figure*}[ht!]
    \centering
    \includegraphics[width=0.495\linewidth]{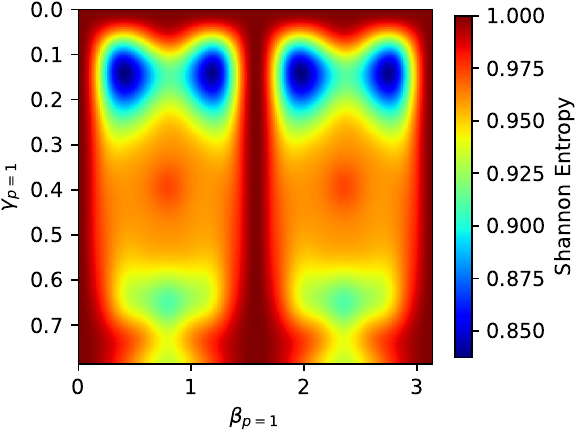}
    \includegraphics[width=0.495\linewidth]{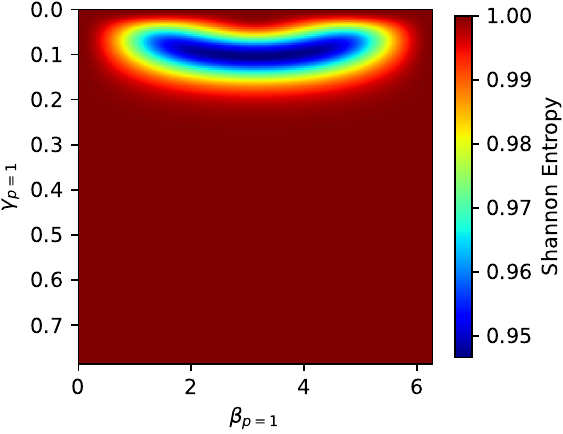}
    \caption{Full QAOA probability distribution Shannon entropy (normalized) $p=1$ angle landscape, for the X mixer (left) and Grover mixer (right). A Shannon entropy of $1$ is the maximum possible value, corresponding to a uniform distribution, and a Shannon entropy of $0$ corresponds to a maximally biased distribution. Recall that $\beta_{p=1}$ here is a QAOA angle parameter, not inverse temperature. }
    \label{fig:Shannon_entropy_heatmap}
\end{figure*}

\begin{figure*}[ht!]
    \centering
    \includegraphics[width=0.495\linewidth]{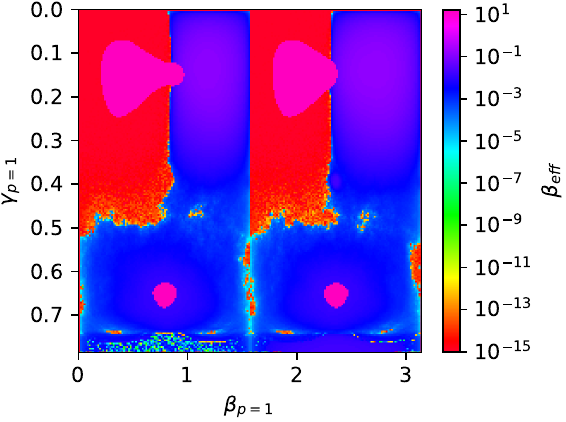}
    \includegraphics[width=0.495\linewidth]{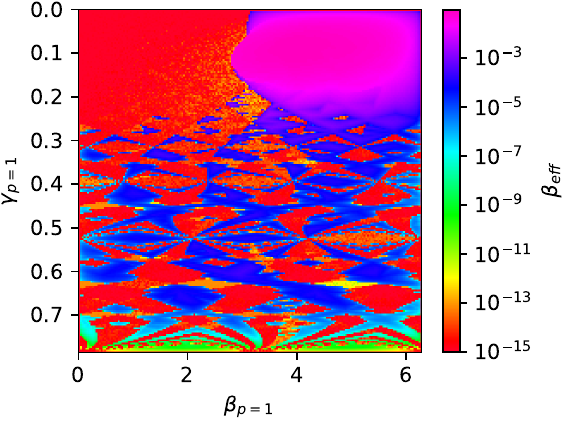}
    \includegraphics[width=0.495\linewidth]{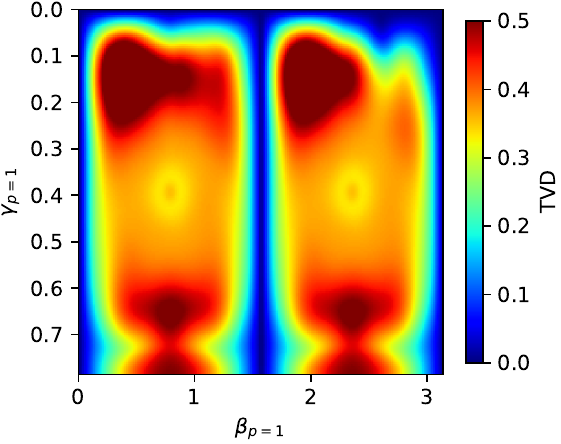}
    \includegraphics[width=0.495\linewidth]{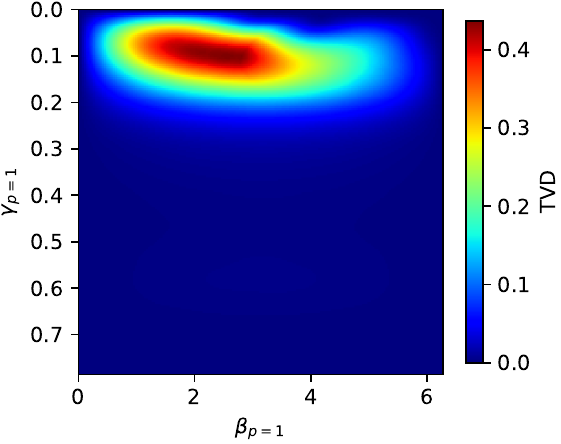}
    \caption{Boltzmann distribution sampling properties of QAOA using the X mixer (left column) and the Grover mixer (right column). Top row contains heatmaps of the best (lowest-error) effective inverse temperature $\beta$ that was fitted to the full noiseless QAOA probability distribution, with a logarithmic scale heatmap. The bottom row contains heatmaps of the error measure, TVD, between the ideal Boltzmann distribution and the QAOA probability distribution, at the best fitted $\beta$. The TVD values, and the corresponding $\beta$ value, are the absolute best-fit that was found -- in other words, the lowest TVD that was found during the numerical fitting process. The heatmap landscape coordinates are given by the depth one QAOA parameters of $\vec{\gamma}$ (y-axis) and $\vec{\beta}$ (x-axis). The type of sampling that would be most interesting computationally is low temperature sampling, so maximize $\beta$. However, we also want accurate sampling, so small TVD measures -- most of the low temperature sampling regimes we see in these heatmaps also have high error rates. 
    Recall that $\beta_{p=1}$ here is a QAOA angle parameter, not inverse temperature. Whereas, $\beta_{\text{eff}}$ (top-row) is the (lowest-error fitted) inverse temperature. 
    Note that the plots in the top row and the bottom row of this figure use different color scales; in the top row red corresponds to very small $\beta_{\text{eff}}$, whereas in the bottom row red corresponds to very high error rates. 
    }
    \label{fig:TVD_Boltzmann_distribution}
\end{figure*}

TVD is bounded to be between $0$ ($P(x)$ and $Q(x)$ are identical), and $1$ means the distributions are disjoint and have no overlap. In practice, a TVD of $0.5$ is the effective maximum error that we will see between two probability distributions. TVD is the error measure which we will aim to minimize in the thermal distribution fitting process. There are many potential error measures to compare two probability distributions; different error measures may produce different response landscapes. Here we use TVD because it is a natural measure, and low error corresponds to high agreement between the two distributions, however there are many other physically meaningful error measures that could be used such as Kullback–Leibler divergence.

The temperature distribution fitting procedure aims to find the Boltzmann distribution (specifically the inverse temperatures $\beta$) that is most closely approximated by the given QAOA sample distribution (regardless of what the temperature is), which is defined by a probability for each of the $2^N$ configurations of the Ising model. The full Boltzmann distribution fitting process involves two primary stages. The ultimate goal is to simply collect a wide range of data-points on the error rate, measured as TVD, across many inverse temperatures $\beta$. The relevant quantity we are aiming to find is which $\beta$ results in the lowest TVD error rate. The first is applying black-box minimizers from scipy~\cite{2020SciPy-NMeth}, namely \texttt{nelder-mead}~\cite{10.1007/s10589-010-9329-3}, \texttt{cobyla}, \texttt{l-bfgs-b}, \texttt{powell}, \texttt{slsqp}~\cite{nocedal2006numerical}, \texttt{trust-constr}~\cite{conn2000trust}, and \texttt{tnc}, using $28$ different starting initial $\beta$ guesses ranging from $10^{5}$ to $10^{-8}$. The second stage performs several grid-searches. The first is over 100 linearly spaced, in logarithmic 10 space, $\beta$ values from $10^{-3}$ to $10^{-15}$. The second is over $\beta$ in steps of $10^{-4}$, starting at $10^{-4}$, until division by zero errors are encountered at sufficiently large $\beta$, at which point the grid-search is terminated. All TVD and $\beta$ estimates are rounded to a reasonable numerical precision of $15$ decimal places. Frequently, this fitting results in precision limits and division by zero errors at very low (attempted) temperature fits, all of which are discarded during this fitting process. From all of these data points, we record the minimum error rate found, and what inverse temperature $\beta$ resulted in that lowest error rate -- in other words, finding what $\beta$ is the closest fit to the given QAOA probability distribution. This procedure is then repeated, independently, for each QAOA setting and mixer configuration. In this way, the fitting procedure is a search over many different probability distributions, at each step computing the error (TVD) between that thermal probability distribution, and the distribution produced by QAOA. The use of the statevector quantum circuit simulation eliminates finite sampling effects as a consideration that must be made when determining whether the QAOA probability distribution approximates a Boltzmann distribution or not. In particular, this inverse temperature fitting may result in many possible $\beta$ values that have the same TVD error rate, up to numerical TVD precision -- however, this never results in high variance of the estimated $\beta$ for low temperatures. At high temperatures, the TVD can plateau, in which case the geometric mean of the computed $\beta$ estimates at that same error rate is used. Appendix~\ref{section:Temperature_fitting_search_landscape} presents some example TVD and $\beta_{\text{eff}}$ numerical fitting plots.

The full Boltzmann distribution sampling requires a significant amount of fitting and therefore computation time. To this end, we use the Python 3 libraries Numba and Numpy~\cite{10.1145/2833157.2833162, harris2020array} in order to significantly speed up these calculations, in particular the calculation of the thermal probability distributions and then the TVD error measures between two probability distributions. 

Lastly, in order to quantify how uniform the QAOA probability distributions are, we turn to the Shannon entropy measure~\cite{shannon1948mathematical} 

\begin{equation}
S(x) = - \sum_{x} p(x) \log_{2^N} p(x),
\end{equation}

where $S(x)$ is the normalized entropy ($1$ means a uniform distribution, and $0$ means a distribution where all probabilities are on one configuration). The logarithm base is equal to the total number of configurations, which is $2^N$ (this gives the measure being normalized), and $p(x)$ is the (noiseless) QAOA probability of that particular $z$ spin configuration. This information theoretic measure is another way of viewing the characteristics of these QAOA distributions; a high Shannon entropy measure across all configurations corresponds to effectively random sampling, and a lower Shannon entropy could indicate good thermal sampling (or, biased and therefore inaccurate thermal sampling).

\begin{figure*}[ht!]
    \centering
    \includegraphics[width=0.495\linewidth]{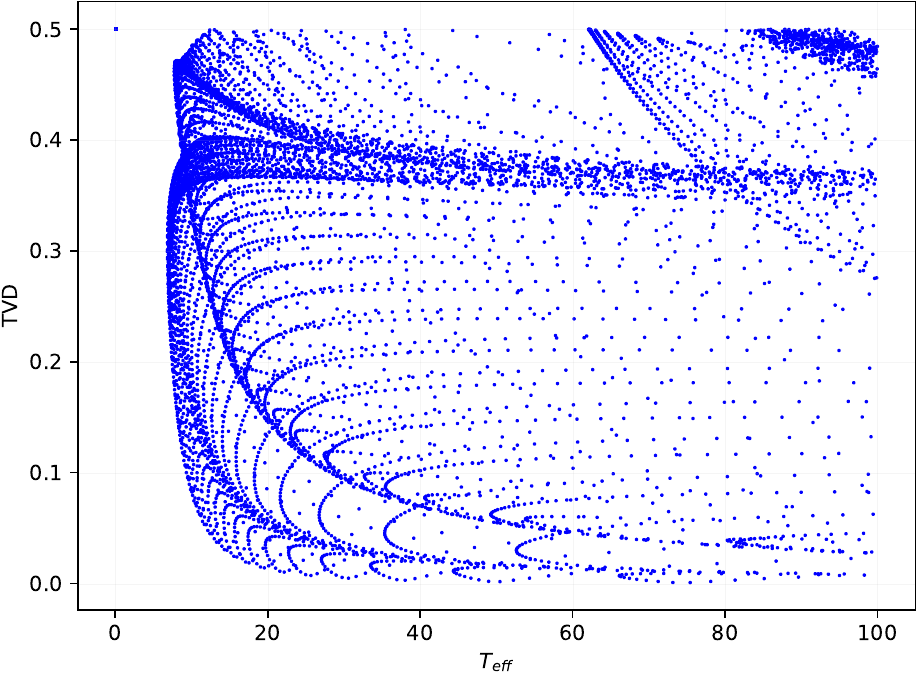}
    \includegraphics[width=0.495\linewidth]{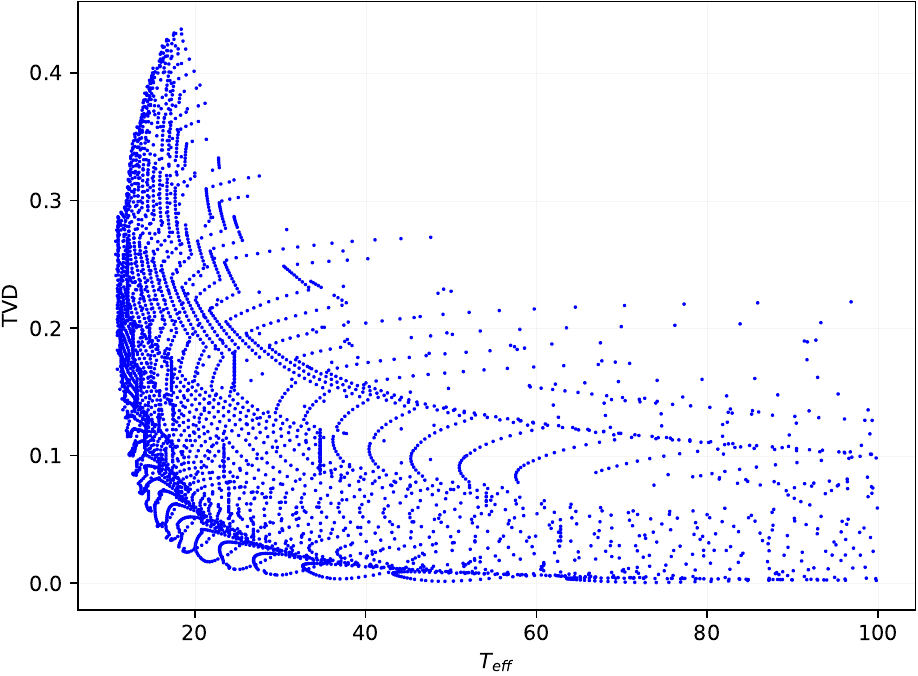}
    \caption{Scatterplot of the TVD error (y-axis), with respect to the ideal Boltzmann distribution, of the best-fitted temperature ($1/\beta_{\text{eff}}$) found for the given QAOA sample distribution, as a function of that effective temperature (x-axis) that was fitted. X mixer QAOA distribution is shown on the left, and the Grover mixer QAOA distribution is shown on the right. TVD closer to $0$ means a better approximation of the Boltzmann distribution at that effective temperature ($T_\text{eff}$). For these tradeoff plots, the aim is to focus on low temperature sampling, and therefore effective temperatures greater than $100$ are not plotted. Specifically, this data is a different representation of the high resolution $200\times 200$ grid-search plots in Fig.~\ref{fig:TVD_Boltzmann_distribution}, but only the points with an estimated effective temperature less than $100$ are shown. We see that there are many $\vec{\beta},\vec{\gamma}$ configurations which result in extremely high error rates of $\approx 0.5$, and for low error rates the effective temperature becomes higher.  }
    \label{fig:scatterplots_X_GM_TVD_temp_tradeoff}
\end{figure*}

\begin{figure}[ht!]
    \centering
    \includegraphics[width=0.999\linewidth]{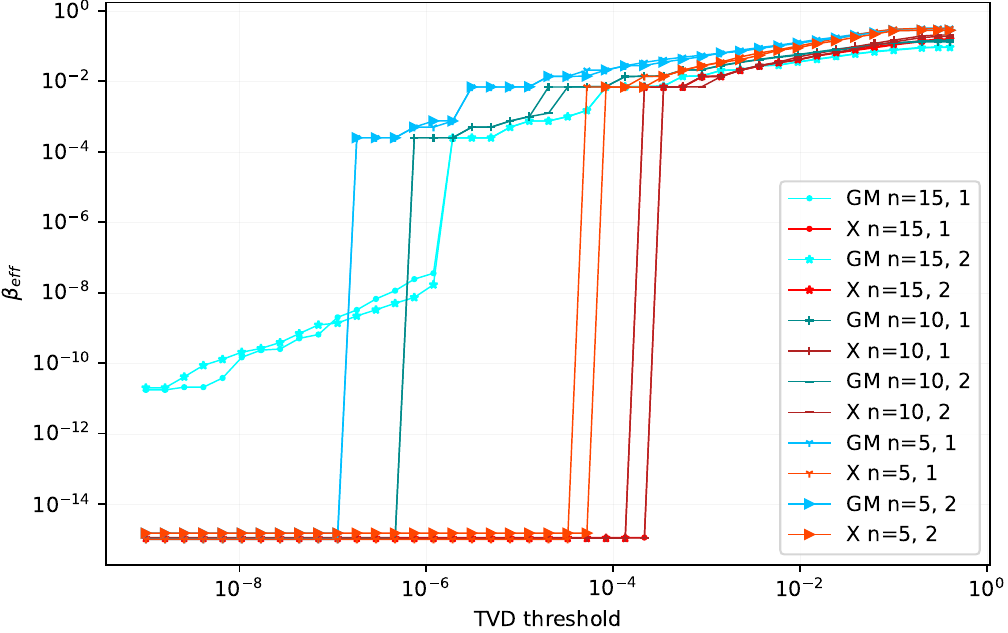}
    \caption{Comparing the two mixers by a TVD threshold analysis where the largest effective fitted $\beta$ (y-axis) that has TVD below the threshold (x-axis) is plotted. Logarithmic scale on both axes. We see a sharp dropoff of the X mixer sampling capability in the effective $\beta$ as the TVD threshold is made smaller, whereas at smaller TVD thresholds Grover mixer QAOA is able to sample at lower temperatures compared to X mixer QAOA. The legend denotes data from $6$ unique SK model instances, where the integers in the legend denote instance 1 and instance 2, with different cost function polynomial coefficients, for each of the three Ising model sizes ($5, 10, 15$ spins). Note that here the y-axis is the inverse temperature $\beta$, not the QAOA angle parameter.   }
    \label{fig:TVD_threshold_mixer_comparison}
\end{figure}

\begin{table}[ht!]
    \begin{center}
        \begin{tabular}{|l|l|l|l|}
            \hline
            Mixer, TVD error threshold & $T_{\text{eff}}$ & $\vec{\beta_{p=1}}$ & $\vec{\gamma_{p=1}}$ \\
            \hline
            $X$ mixer, TVD $0.1$ & $9.32$ & $2.6522$ & $0.0710$ \\
            \hline
            $X$ mixer, TVD $0.01$ & $24.98$ & $2.6206$ & $0.0237$ \\
            \hline
            $X$ mixer, TVD $0.001$ & $73.16$ & $2.6206$ & $0.007893$ \\
            \hline
            \hline
            \hline
            GM mixer, TVD $0.1$ & $12.75$ & $4.1993$ & $0.05131$ \\
            \hline
            GM mixer, TVD $0.01$ & $28.12$ & $4.3256$ & $0.01973$ \\
            \hline
            GM mixer, TVD $0.001$ & $70.39$ & $4.2625$ & $0.007893$ \\
            \hline
        \end{tabular}
    \end{center}
    \caption{ Summary of some QAOA Boltzmann distribution sampling statistics, at three different TVD error rate thresholds, along with the QAOA angles that result in the thermal distribution. Data from one of the $15$-spin SK models.   }
    \label{table:results_summary}
\end{table}

\section{Results}
\label{section:results}

Fig.~\ref{fig:QAOA_energy_heatmap} presents a $200\times 200$ QAOA energy landscape grid of $\beta_{p=1}$ and $\gamma_{p=1}$ parameters. This is where we start the thermal sampling investigation, by first simply checking the expectation value landscape. Importantly, these energies are far away from both the global minimum and maximum, so we do not expect to be sampling very low temperatures (which is reasonable, typically larger $p$ step sizes are required in order to get to lower temperatures). Much of the angle search landscape is quite flat. The angle symmetries used here are a more focused and higher resolution gridsearch than the full angle periodicities -- see Appendix~\ref{section:appendix_QAOA_energy_landscape}. Next, Fig.~\ref{fig:Shannon_entropy_heatmap} shows the Shannon entropy of the QAOA probability distributions, which show that the entropy is symmetric regardless of the expectation value (comparing to Fig.~\ref{fig:QAOA_energy_heatmap}), but also does not deviate very significantly from a nearly uniform entropy (the minimum possible entropy measure here is $0$). Moreover, GM-QAOA has a higher overall configuration Shannon entropy than X mixer QAOA. Shannon entropy smaller than $1$ means the measured probability distribution is biased, which we may expect would produce good thermal states. However, what we also want is low temperature thermal state sampling (e.g the blue regions in Fig.~\ref{fig:QAOA_energy_heatmap}), which is also preparing an accurate thermal state. This means that any ``good'' (meaning, accurate and also low-temperature) Boltzmann states in the variational QAOA $p=1$ landscape are somewhere near or within the blue/light blue regions of Fig.~\ref{fig:QAOA_energy_heatmap}, and also within the blue/light blue regions of Fig.~\ref{fig:Shannon_entropy_heatmap}, but without having biased sampling of energy levels (which is what GM-QAOA naturally provides, X-QAOA does not necessarily provide this guarantee).

Next we turn to the fitted Boltzmann distribution data, summarized in Fig.~\ref{fig:TVD_Boltzmann_distribution}. This figure shows side by side both the effective fitted inverse temperatures, and the corresponding lowest error rates that were found in this fitting process. The log-scale heatmap of $\beta_{\text{eff}}$ in Fig.~\ref{fig:TVD_Boltzmann_distribution} reveals a highly complex landscape, and for GM-QAOA (upper-right panel) this landscape qualitatively shows some self-similarity regions. The self-similarity pattern emerging in the GM-QAOA case suggests that it is connected to the global nature of the mixer. The mechanism causing this self similarity is not clear particularly because this is a log-scale closest-$\beta$ landscape (not, an exact thermal sampling at $\beta$ landscape), and this is left as a future research topic. Comparing to TVD for X mixer QAOA, it is clear that in the regions where $\beta_{\text{eff}}$ is large, the corresponding error rate is also maximized to $0.5$. This is important because this shows that the regions where the energy of the Hamiltonian is strictly minimized for X mixer QAOA, the probability distributions are completely inaccurate Boltzmann distributions for that low temperature. In the case of the X-mixer (left-column), the best-fit temperature is very low (which is good because that is the most computationally challenging region in general), and encompasses a decent region of the angle parameter space (magenta color), but also has the highest possible error rate (TVD of $\approx 0.5$), meaning this is a very bad thermal distribution. Most of the GM-QAOA angle parameter space (right-column in Fig.~\ref{fig:TVD_Boltzmann_distribution}) has a very low error rate, but also a very high temperature distribution. There are some small regions of Fig.~\ref{fig:TVD_Boltzmann_distribution} that have both a reasonably small error rate but also a low-temperature distribution.

To more closely examine the low-error rate, low-temperature sampling region of the parameter space seen in Fig.~\ref{fig:TVD_Boltzmann_distribution}, next Fig.~\ref{fig:scatterplots_X_GM_TVD_temp_tradeoff} shows the tradeoff between the best fitted effective $\beta$ and the TVD to that ideal Boltzmann distribution, for specific low temperature points from these QAOA angle gridsearches. This type of representation of the data is necessary because Fig.~\ref{fig:TVD_Boltzmann_distribution} does not contain large regions of both low TVD and high temperature -- most of the regions are either very high error (TVD $\approx 0.5$) or high temperature. In Fig.~\ref{fig:scatterplots_X_GM_TVD_temp_tradeoff} we see the tradeoff illustrated clearly - where for these depth one QAOA circuits, there are no very low temperature and also low error rate QAOA probability distributions. To this end, Table~\ref{table:results_summary} details the lowest temperature distributions that were found, under three increasingly small TVD thresholds. The trend is that as this error threshold decreases, so does the (largest) effective $\beta$ that is under that threshold. Fig.~\ref{fig:TVD_threshold_mixer_comparison} then quantifies this in more detail by plotting the lowest temperature (largest $\beta$) as a function of decreasingly small TVD thresholds. Here, in Fig.~\ref{fig:TVD_threshold_mixer_comparison} we extend this analysis to include two different random-coefficient instances for $5$, $10$, and $15$-spin SK models. What we find is that there is a sharp dropoff for the X mixer around $10^{-3}$ TVD accuracy where the best effective $\beta$ is essentially random sampling. In general, GM-QAOA is able to generate more accurate thermal distributions compared to X-mixer QAOA across this set of different SK models. For Grover mixer QAOA on the $15$-spin models, there is a steadier downward trend, such that even for very small TVD thresholds, the $\beta_{\text{eff}}$ is still many orders of magnitude larger than for the X mixer QAOA distributions. At very high TVD, for example at $0.3$, the X mixer variant can sample at lower temperatures than GM-QAOA. This shows that GM-QAOA is a higher accuracy Gibbs distribution sampler compared to X-QAOA, at high temperatures. The reason for this could be because of the entropic properties of GM-QAOA, providing an entirely unbiased probability distribution for degenerate configurations in the same energy level. The reason both of the QAOA mixer variants could not achieve temperatures closer to the ground-state is because the evolution was only one-round; future work should therefore pursue studying the properties of higher depth GM-QAOA for thermal sampling. However, due to the exponential scaling that comes with a grid-search at a fixed resolution for higher $p$ depth, alternative methods besides grid-searches will need to be utilized. Empirically, it is clear from Fig.~\ref{fig:TVD_threshold_mixer_comparison} that GM-QAOA, at least for depth $1$, is a better thermal sampler than X mixer QAOA. 

To show the size-dependence, and instance-dependence, of these reported metrics, Appendix~\ref{section:10_spin_results_appendix} reports the same style of plots but on a $10$-spin SK model, and Appendix~\ref{section:5_spin_results_appendix} reports results from a $5$-spin SK model. Appendix~\ref{section:10_spin_results_appendix} and Appendix~\ref{section:5_spin_results_appendix} show that the reported results are quite similar, although not identical, across different Ising models and different problem sizes.

\section{Discussion and Conclusion}
\label{section:conclusion}

We have demonstrated that single round QAOA can generate approximate Boltzmann distributions, and provided clear quantitative tradeoffs between temperature and accuracy, for a high resolution QAOA angle parameter search of $200\times200$. We find that depth one Grover mixer QAOA can prepare more accurate Boltzmann distributions than X mixer QAOA, at relatively high temperatures. As an example, with a TVD error rate of $0.001$ (or better), depth one QAOA can sample the $15$-spin SK model at an effective temperature of $70.39$, whereas for X mixer QAOA the lowest effective temperature that can be sampled with a TVD error rate of $0.001$ or better is $73.16$ (see Table~\ref{table:results_summary}). For both QAOA variants, there is a consistent tradeoff where the best-fitted lower temperature sampling also necessarily means a higher error rate. Very low error rate Boltzmann sampling can be achieved by these two depth one QAOA samplers, but only for higher temperature distributions. We hypothesize that the GM-QAOA thermal sampling properties being more accurate than X-mixer is likely due to the ``fair-sampling'' property guaranteed by GM-QAOA~\cite{Pelofske_2021, Golden_2022_fair, Bartschi_2020} which means that configurations at the same energy level have uniform amplitudes and therefore uniform measurement probabilities. In other words, degeneracy lifting due to the transverse field mixer causes non-thermal distribution preparation, at a configuration level. However, this is not the sole reason for the thermal sampling properties observed in this study the ``variational'' parameters $\vec{\beta}$, $\vec{\gamma}$ inherent to QAOA impact the resulting probability distributions substantially. The role of degeneracy lifting causing non-thermal sampling at a configuration level, as opposed to a density of states analysis, for X mixer QAOA deserves further future study. GM-QAOA provides maximum entropy distributions (uniform distributions) for each energy level of a classical Hamiltonian. This property is required for good thermal sampling, but is not sufficient by itself. The empirical numerical finding that GM-QAOA performs better at this task, at least for shallow depths, should be studied for higher depth QAOA to examine whether this property holds for very-low-temperature sampling, which would be a very relevant quantum computational capability, in particular for frustrated and exotic magnetic systems. Thermal state approximate sampling with relatively short depth quantum algorithms, such as QAOA, could be a good application for near term quantum computer advantage demonstrations. Further study should also be done on applying GM-QAOA thermal sampling to highly frustrated magnetic models, namely systems where traditional Monte Carlo methods begin to fail. Importantly, currently there is no clear algorithm for selecting $\vec{\beta}, \vec{\gamma}$ parameters that do produce thermal states, even in a variational setting -- here we have shown that such parameters do exist, and what their values are for some specific Ising model and mixer combinations. This line of inquiry deserves future study at higher-depth QAOA, focusing on how such parameters can be selected.

The practical considerations of QAOA circuit compilation are not considered in this study. In particular, the circuit depth per layer of GM-QAOA is greater than that of X mixer QAOA~\cite{Pelofske_2021, Golden_2022_fair, Bartschi_2020}, however, this exact complexity scaling is dependent on the underlying quantum computer architecture onto which the circuit would be compiled. Another limitation of this study is that only a total of $6$ Ising models were considered. As discussed in Section~\ref{section:methods}, because of QAOA parameter concentration single instances of Hamiltonians are representative of that class of models. However, precise distribution properties can vary, and quantifying ensemble properties is an important consideration of quantum algorithms. Therefore, future study should address a thorough quantification of the QAOA thermal sampling properties on very large ensembles of problem instances.

One of the key findings of this study is that the highly optimized best performing QAOA angles, in terms of (classical) Hamiltonian energy, are not the best Boltzmann distribution samplers -- in fact, those angles result in probability distributions that are not close to any Boltzmann distribution. This is especially true for the standard X mixer QAOA. Nevertheless, future study should perform a similar type of analysis but for higher depth QAOA circuits, using optimized QAOA angles, to assess what happens to the thermal distribution properties of QAOA in the higher depth regime.

\section*{Data Availability}
\label{section:data}
All data from this study is publicly available as a Zenodo dataset~\cite{pelofske_2025_17316067}.

\section*{Acknowledgments}
\label{section:acknowledgments}
This work was supported by the U.S. Department of Energy through the Los Alamos National Laboratory. Los Alamos National Laboratory is operated by Triad National Security, LLC, for the National Nuclear Security Administration of U.S. Department of Energy (Contract No. 89233218CNA000001). Research presented in this article was supported by the NNSA's Advanced Simulation and Computing Beyond Moore's Law Program at Los Alamos National Laboratory. LA-UR-25-30026.

\appendix

\begin{figure*}[ht!]
    \centering
    \includegraphics[width=0.495\linewidth]{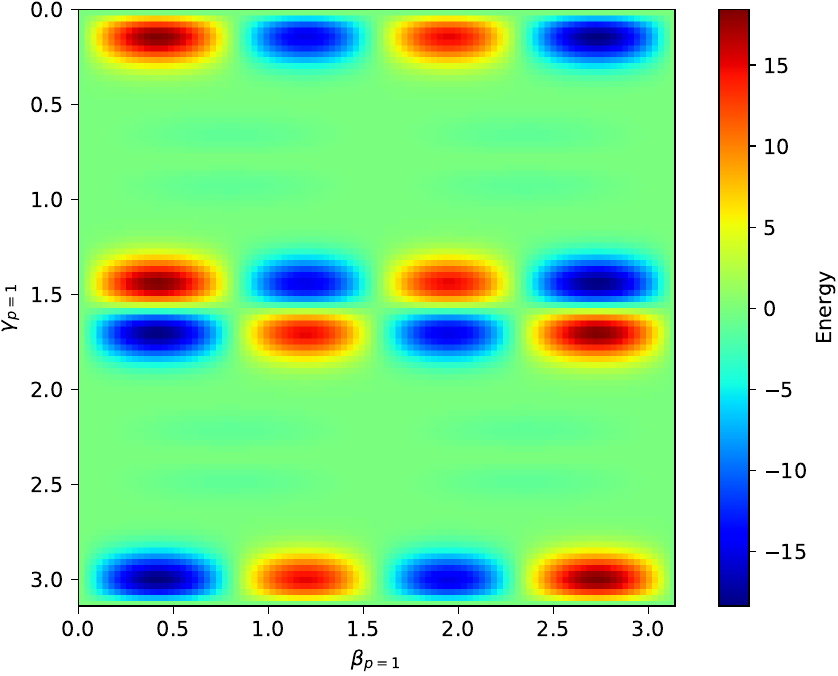}
    \includegraphics[width=0.495\linewidth]{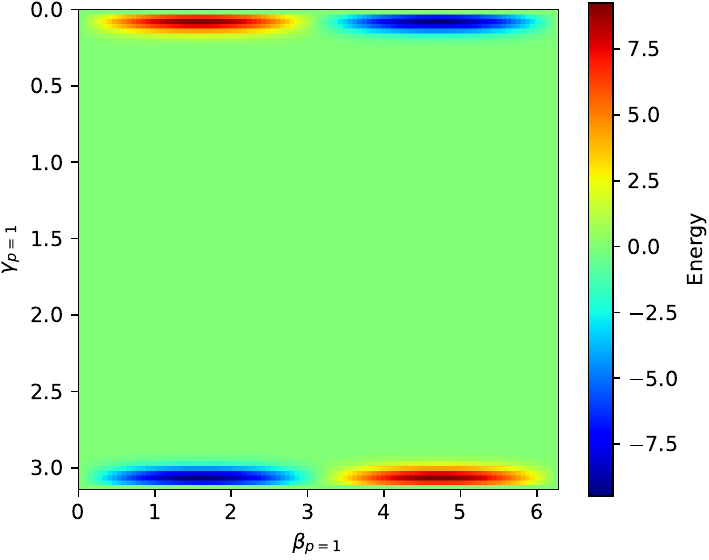}
    \caption{Energy $p=1$ angle landscape for the X mixer (left) and Grover mixer (right). The search landscape here is over the range of the QAOA angle symmetries for this particular SK Ising model instance, for each mixer (notice that the Grover mixer symmetry is from $0$ to $2\pi$). Using these full angle search spaces, we narrow in on smaller regions to examine for all subsequent simulations (specifically, only examining the range from $[0, \frac{\pi}{4}]$ for $\gamma_1$). Data from one of the $15$-spin SK models.  }
    \label{fig:full_angle_gridsearch_energy}
\end{figure*}

\section{Full QAOA $p=1$ Energy Landscapes}
\label{section:appendix_QAOA_energy_landscape}

Fig.~\ref{fig:full_angle_gridsearch_energy} shows $100\times 100$ QAOA angle landscapes, in terms of Hamiltonian energy for the full periodicity of both the phase separator and the mixing Hamiltonian. The aim here was to find what periodicities exist in this angle search range, and in particular what simplifications could be made to the angle search spaces. The result was that we used the full $[0, 2\pi]$ period for the Grover mixer angle range, and the standard $[0, \pi]$ range for the X mixer angle range. The periodicity of the $\gamma$ angle was set to just $[0, \frac{\pi}{4}]$ due to the mirror symmetry we can see in Fig.~\ref{fig:full_angle_gridsearch_energy}, resulting from the mirror symmetry of the Ising model. Fig.~\ref{fig:QAOA_energy_heatmap} shows the more focused QAOA energy landscape, which was then used in all subsequent QAOA simulations.

\section{Example Boltzmann Distribution Fitting Search Landscapes}
\label{section:Temperature_fitting_search_landscape}

\begin{figure*}[ht!]
    \centering
    \includegraphics[width=0.32\linewidth]{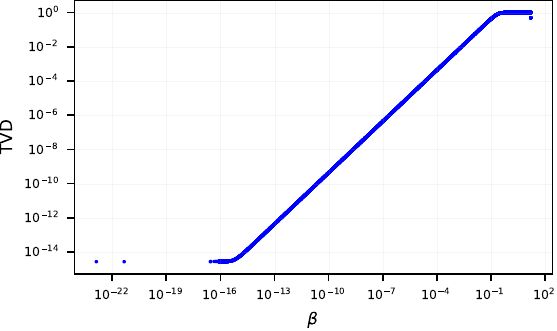}
    \includegraphics[width=0.32\linewidth]{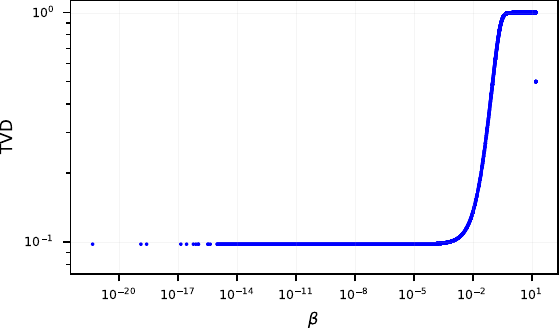}
    \includegraphics[width=0.32\linewidth]{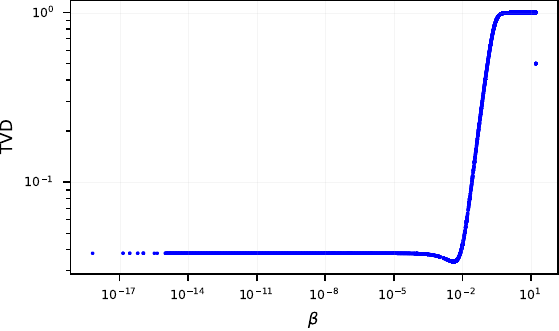}
    \includegraphics[width=0.32\linewidth]{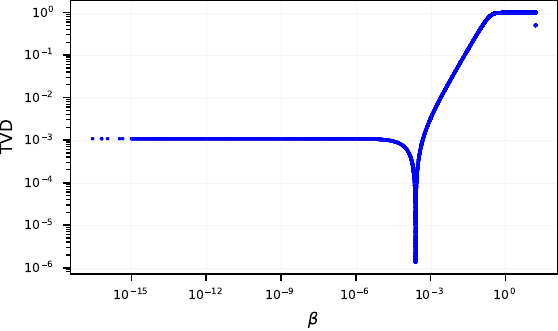}
    \includegraphics[width=0.32\linewidth]{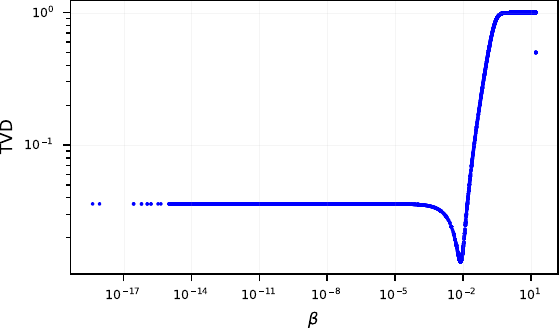}
    \includegraphics[width=0.32\linewidth]{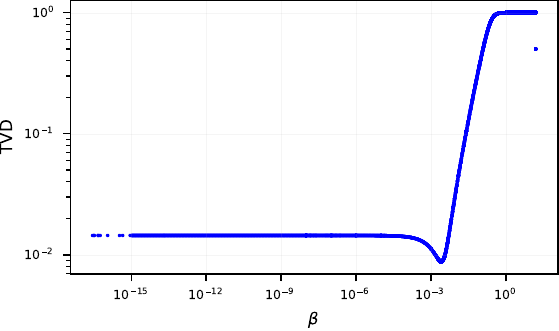}
    \caption{ A small set of examples of the error (TVD) vs $\beta$ Boltzmann sampling response surfaces, for different mixers and $\vec{\beta},\vec{\gamma}$ combinations. Log scale y-axis and x-axis. $15$-qubit SK model. Note that here $\beta$ is inverse temperature, not the QAOA mixer parameter.   }
    \label{fig:example_TVD_beta_response_surfaces}
\end{figure*}

Fig.~\ref{fig:example_TVD_beta_response_surfaces} reports example $\beta$ vs TVD thermal distribution plots, for selected points in the $\vec{\beta},\vec{\gamma}$ search space for both Grover and X mixer QAOA numerical simulations, on one of the $15$-qubit SK models. Each point in each subplot corresponds to a single objective function call that was performed, either in a gridsearch or a black-box optimizer. This shows that not only is the fitting process for each distribution high-resolution, but also that there does exist a unique minimum TVD typically -- or, in some cases the distribution is very high temperature and the TVD plateaus.

\section{$10$-Spin QAOA Thermal Sampling Results}
\label{section:10_spin_results_appendix}

\begin{figure*}[ht!]
    \centering
    \includegraphics[width=0.495\linewidth]{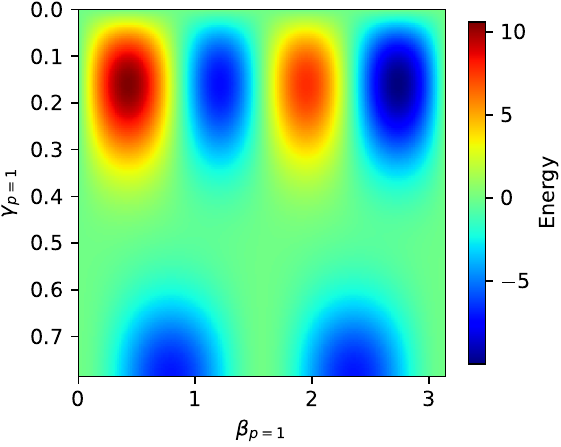}
    \includegraphics[width=0.495\linewidth]{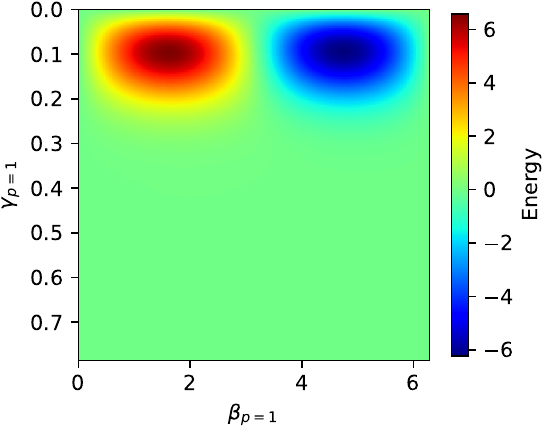}
    \caption{ Hamiltonian energy as a function of both $\vec{\beta}$ and $\vec{\gamma}$ for a $10$-spin SK model, for the X mixer (left) and Grover mixer (right). Same format as Fig.~\ref{fig:QAOA_energy_heatmap}.   }
    \label{fig:QAOA_energy_heatmap_10_spin}
\end{figure*}

\begin{figure*}[ht!]
    \centering
    \includegraphics[width=0.495\linewidth]{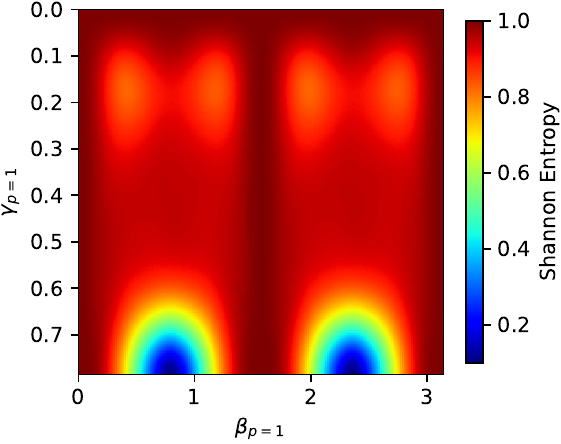}
    \includegraphics[width=0.495\linewidth]{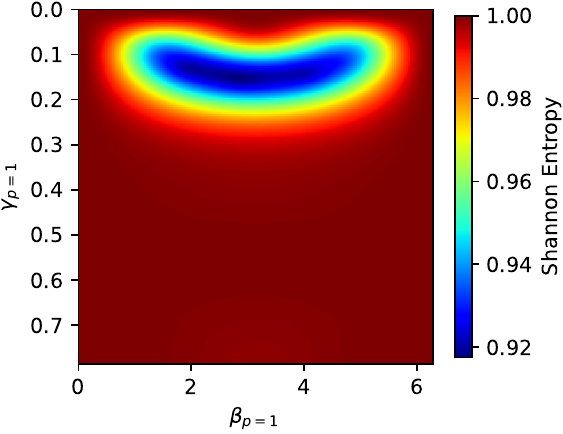}
    \caption{ Shannon entropy of the full QAOA probability distribution from a $10$-spin SK model for the X mixer (left) and Grover mixer (right). Same format as Fig.~\ref{fig:Shannon_entropy_heatmap}.  }
    \label{fig:Shannon_entropy_heatmap_10_spin}
\end{figure*}

\begin{figure*}[ht!]
    \centering
    \includegraphics[width=0.495\linewidth]{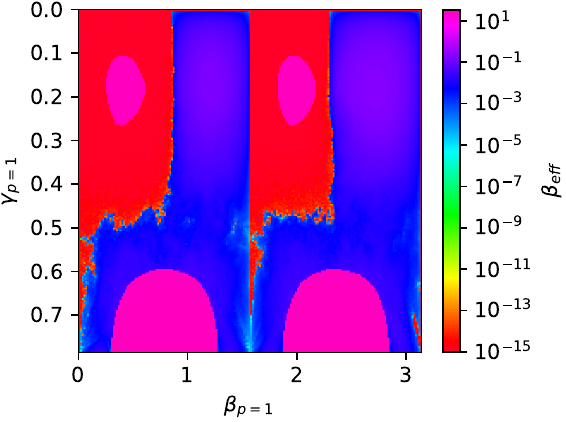}
    \includegraphics[width=0.495\linewidth]{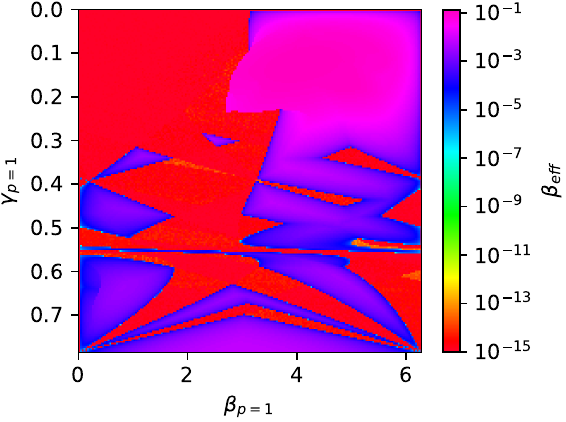}
    \includegraphics[width=0.495\linewidth]{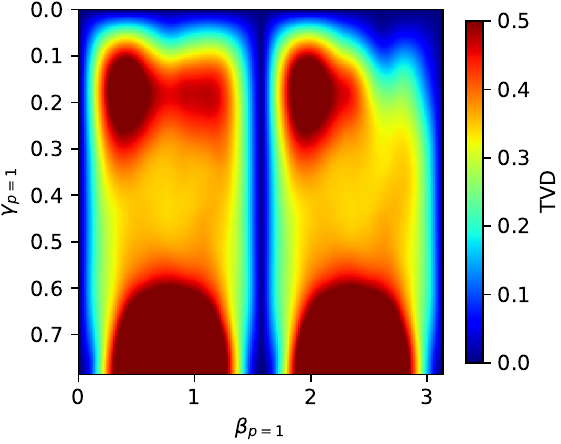}
    \includegraphics[width=0.495\linewidth]{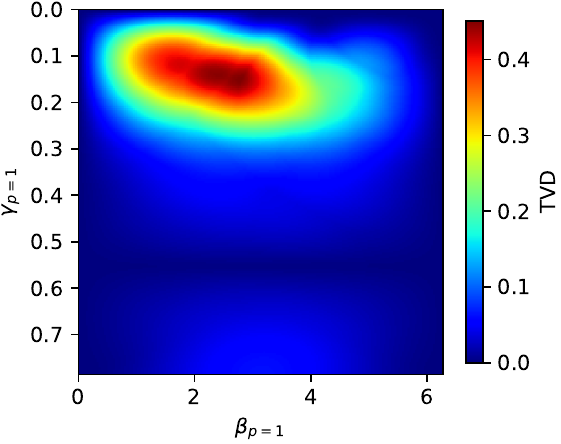}
    \caption{ Boltzmann sampling analysis with respect to both TVD and the best-fitted $\beta$ for X mixer (left) and Grover mixer QAOA (right). Data from a $10$-spin SK model, and the plot format is the same as Fig.~\ref{fig:TVD_Boltzmann_distribution}.  }
    \label{fig:TVD_Boltzmann_distribution_10_spin}
\end{figure*}

\begin{figure*}[ht!]
    \centering
    \includegraphics[width=0.495\linewidth]{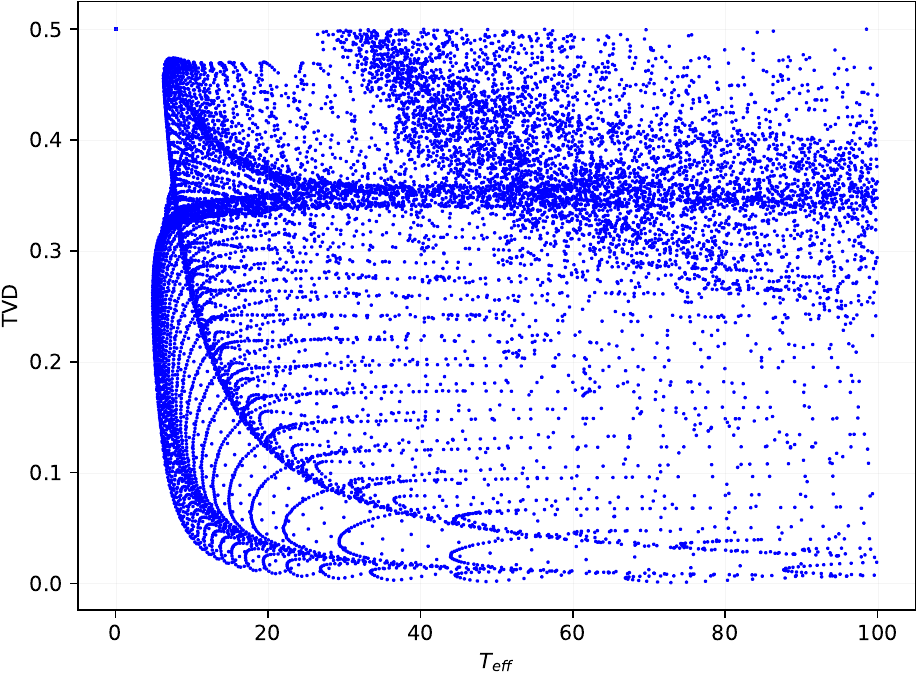}
    \includegraphics[width=0.495\linewidth]{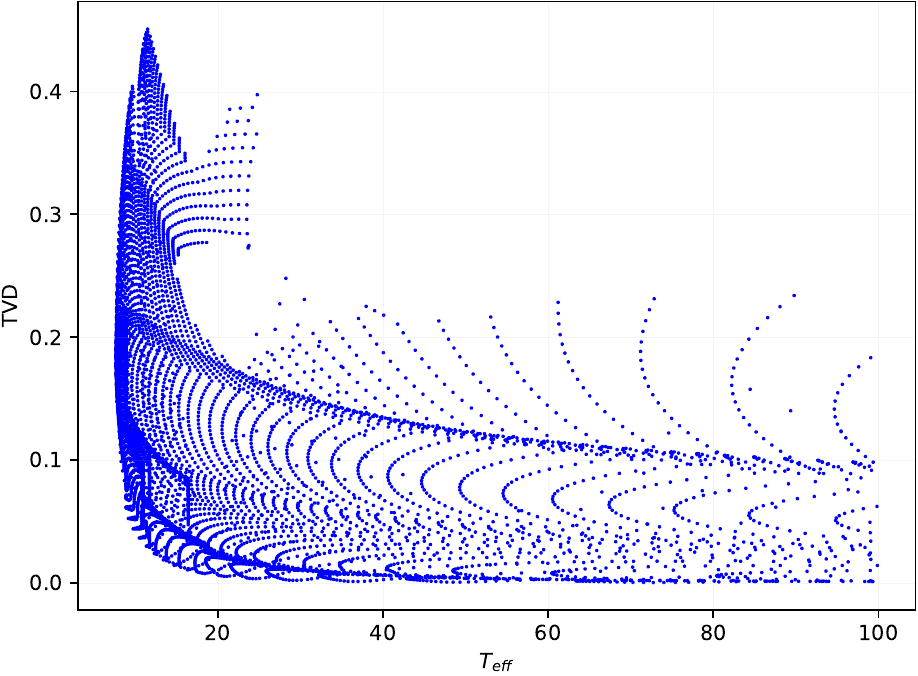}
    \caption{ Scatterplot showing the low-temperature and error (TVD) tradeoff distribution for X mixer QAOA (left) and GM-QAOA (right) on a $10$-spin SK model. The plot format is the same as Fig.~\ref{fig:scatterplots_X_GM_TVD_temp_tradeoff}.  }
    \label{fig:scatterplots_X_GM_TVD_temp_tradeoff_10_spin}
\end{figure*}

The main text results focus on a $15$-spin SK model, but here we present the same analysis but on a $10$-spin SK model. Fig.~\ref{fig:QAOA_energy_heatmap_10_spin} shows the Hamiltonian expectation value, Fig.~\ref{fig:Shannon_entropy_heatmap_10_spin} plots the distribution Shannon entropy, Fig.~\ref{fig:TVD_Boltzmann_distribution_10_spin} plots the Boltzmann distribution temperature and TVD as heatmaps, and Fig.~\ref{fig:scatterplots_X_GM_TVD_temp_tradeoff_10_spin} plots the tradeoff between temperature and error rate as scatterplots for low-temperature sampling. Like in Fig.~\ref{fig:TVD_Boltzmann_distribution}, the GM-QAOA distribution fitted-temperature landscape (top-right) in Fig.~\ref{fig:TVD_Boltzmann_distribution_10_spin} has an interesting partial self-similarity pattern. 

\section{$5$-Spin QAOA Thermal Sampling Results}
\label{section:5_spin_results_appendix}

\begin{figure*}[ht!]
    \centering
    \includegraphics[width=0.495\linewidth]{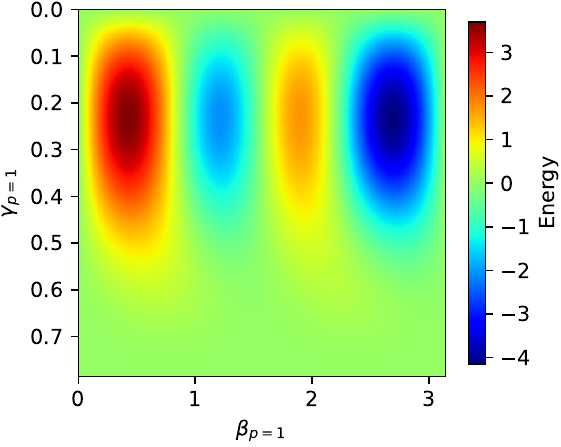}
    \includegraphics[width=0.495\linewidth]{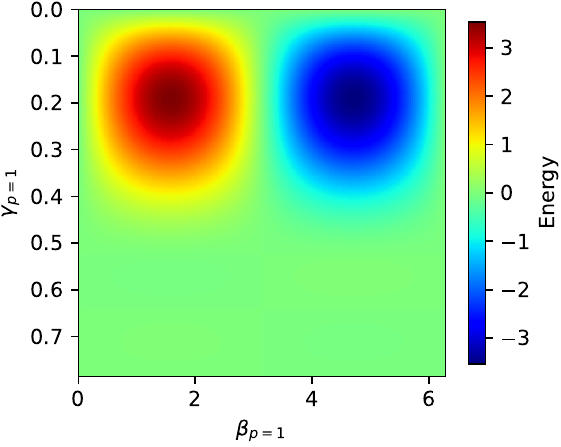}
    \caption{ Hamiltonian energy as a function of both $\vec{\beta}$ and $\vec{\gamma}$ for a $5$-spin SK model, for the X mixer (left) and Grover mixer (right). Same format as Fig.~\ref{fig:QAOA_energy_heatmap}.  }
    \label{fig:QAOA_energy_heatmap_5_spin}
\end{figure*}

\begin{figure*}[ht!]
    \centering
    \includegraphics[width=0.495\linewidth]{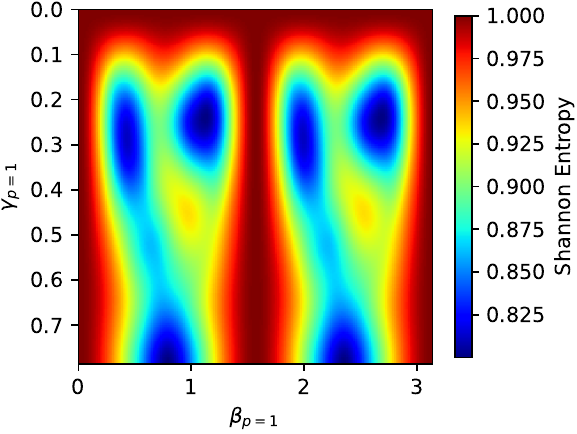}
    \includegraphics[width=0.495\linewidth]{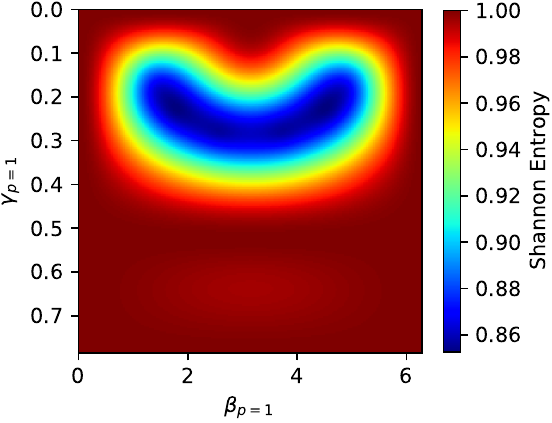}
    \caption{ Shannon entropy of the full QAOA probability distribution from a $5$-spin SK model for the X mixer (left) and Grover mixer (right).  Same format as Fig.~\ref{fig:Shannon_entropy_heatmap}. }
    \label{fig:Shannon_entropy_heatmap_5_spin}
\end{figure*}

\begin{figure*}[ht!]
    \centering
    \includegraphics[width=0.495\linewidth]{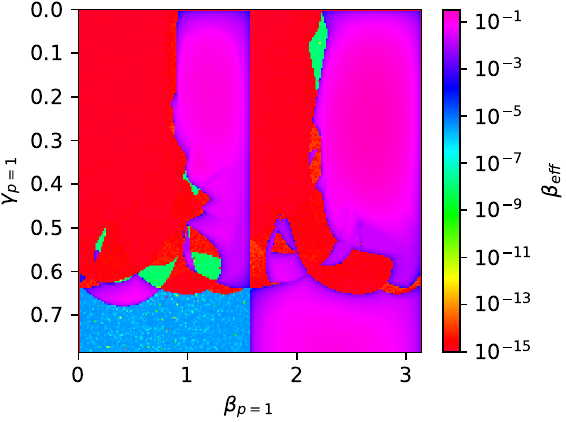}
    \includegraphics[width=0.495\linewidth]{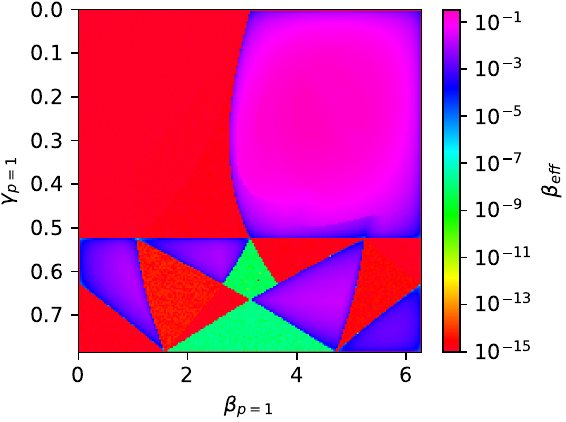}
    \includegraphics[width=0.495\linewidth]{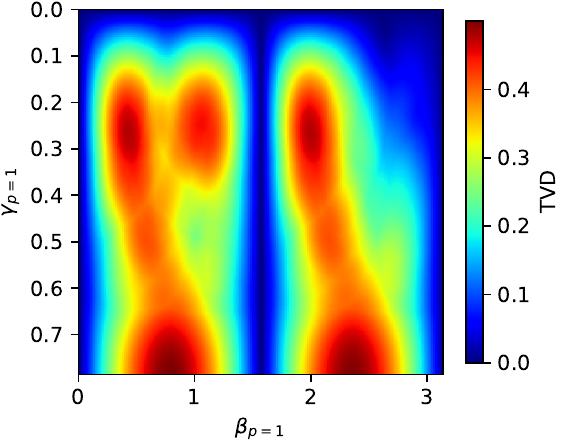}
    \includegraphics[width=0.495\linewidth]{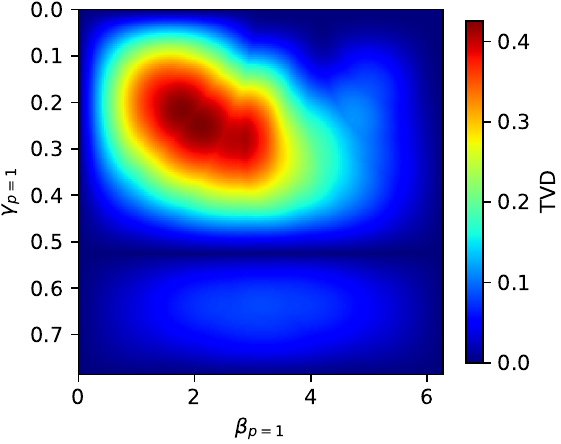}
    \caption{ Boltzmann sampling analysis with respect to both TVD and the best-fitted $\beta$ for X mixer (left) and Grover mixer QAOA (right). Data from a $5$-spin SK model, and the plot format is the same as Fig.~\ref{fig:TVD_Boltzmann_distribution}. }
    \label{fig:TVD_Boltzmann_distribution_5_spin}
\end{figure*}

\begin{figure*}[ht!]
    \centering
    \includegraphics[width=0.495\linewidth]{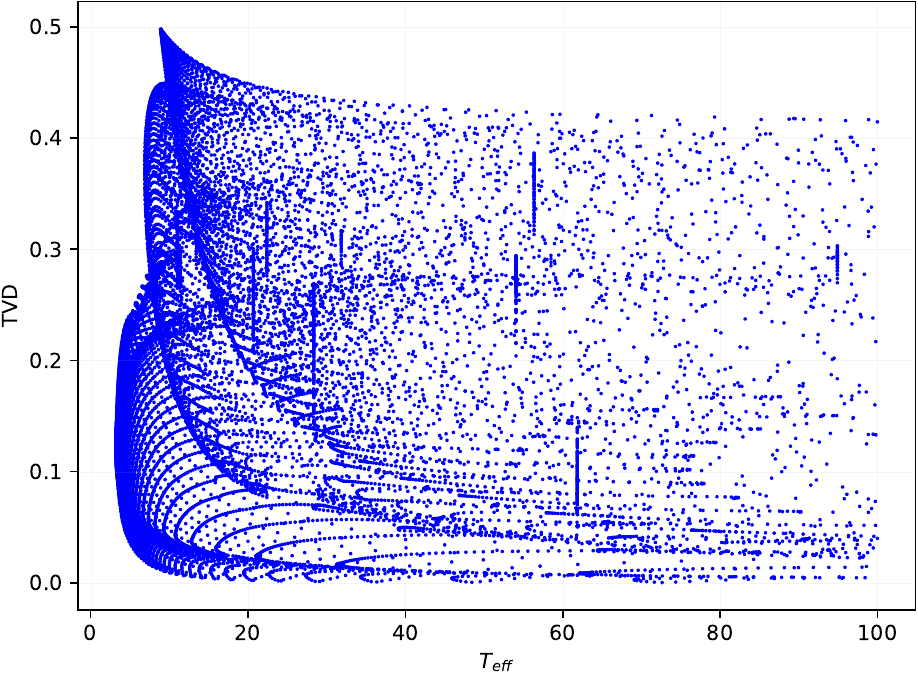}
    \includegraphics[width=0.495\linewidth]{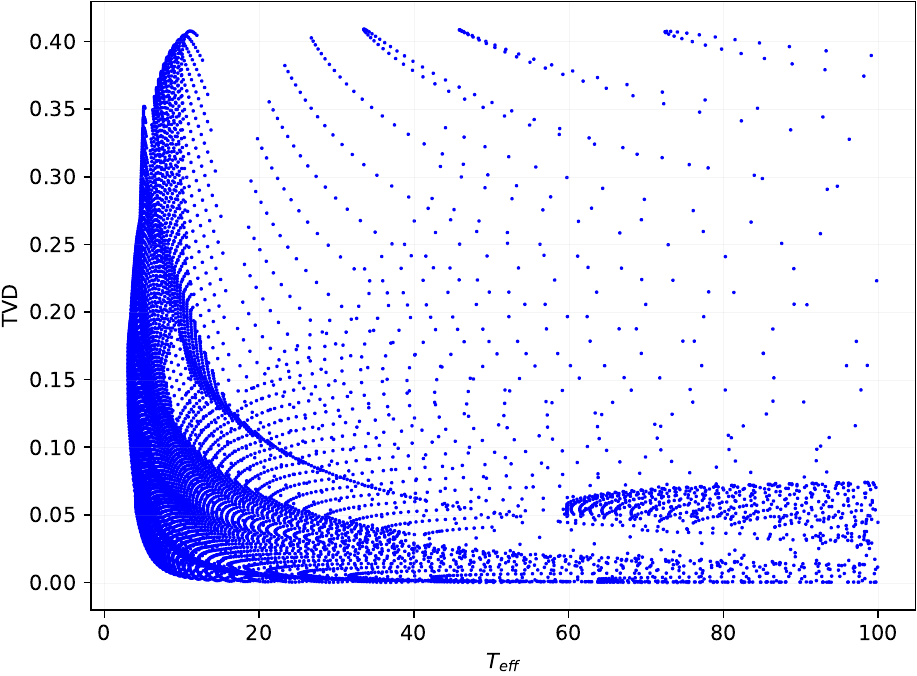}
    \caption{ Scatterplot showing the low-temperature and error (TVD) tradeoff distribution for X mixer QAOA (left) and GM-QAOA (right) on a $5$-spin SK model. The plot format is the same as Fig.~\ref{fig:scatterplots_X_GM_TVD_temp_tradeoff}. }
    \label{fig:scatterplots_X_GM_TVD_temp_tradeoff_5_spin}
\end{figure*}

The main text results focus on a $15$-spin SK model, but here we present the same analysis but on a $5$-spin SK model. Fig.~\ref{fig:QAOA_energy_heatmap_5_spin} shows the Hamiltonian expectation value, Fig.~\ref{fig:Shannon_entropy_heatmap_5_spin} plots the distribution Shannon entropy, Fig.~\ref{fig:TVD_Boltzmann_distribution_5_spin} plots the Boltzmann distribution temperature and TVD as heatmaps, and Fig.~\ref{fig:scatterplots_X_GM_TVD_temp_tradeoff_5_spin} plots the tradeoff between temperature and error rate as scatterplots for low-temperature sampling. Like in Fig.~\ref{fig:TVD_Boltzmann_distribution}, the GM-QAOA distribution fitted-temperature landscape (top-right) in Fig.~\ref{fig:TVD_Boltzmann_distribution_5_spin} has an interesting partial self-similarity pattern. Combined, these $5$-spin SK model results, the $10$-spin SK model results (Appendix~\ref{section:10_spin_results_appendix}), and the $15$-spin SK model results, all show empirically that with respect to the effective thermal sampling properties, QAOA landscapes do concentrate (e.g., are similar) for both mixers, just as they do for the cost function landscapes (such as in Fig.~\ref{fig:QAOA_energy_heatmap_5_spin}, Fig.~\ref{fig:QAOA_energy_heatmap_10_spin}, Fig.~\ref{fig:QAOA_energy_heatmap}).

\clearpage

\bibliographystyle{apsrev4-2-titles}
\bibliography{references}

\begin{thebibliography}{87}%
\makeatletter
\providecommand \@ifxundefined [1]{%
 \@ifx{#1\undefined}
}%
\providecommand \@ifnum [1]{%
 \ifnum #1\expandafter \@firstoftwo
 \else \expandafter \@secondoftwo
 \fi
}%
\providecommand \@ifx [1]{%
 \ifx #1\expandafter \@firstoftwo
 \else \expandafter \@secondoftwo
 \fi
}%
\providecommand \natexlab [1]{#1}%
\providecommand \enquote  [1]{``#1''}%
\providecommand \bibnamefont  [1]{#1}%
\providecommand \bibfnamefont [1]{#1}%
\providecommand \citenamefont [1]{#1}%
\providecommand \href@noop [0]{\@secondoftwo}%
\providecommand \href [0]{\begingroup \@sanitize@url \@href}%
\providecommand \@href[1]{\@@startlink{#1}\@@href}%
\providecommand \@@href[1]{\endgroup#1\@@endlink}%
\providecommand \@sanitize@url [0]{\catcode `\\12\catcode `\$12\catcode `\&12\catcode `\#12\catcode `\^12\catcode `\_12\catcode `\%12\relax}%
\providecommand \@@startlink[1]{}%
\providecommand \@@endlink[0]{}%
\providecommand \url  [0]{\begingroup\@sanitize@url \@url }%
\providecommand \@url [1]{\endgroup\@href {#1}{\urlprefix }}%
\providecommand \urlprefix  [0]{URL }%
\providecommand \Eprint [0]{\href }%
\providecommand \doibase [0]{https://doi.org/}%
\providecommand \selectlanguage [0]{\@gobble}%
\providecommand \bibinfo  [0]{\@secondoftwo}%
\providecommand \bibfield  [0]{\@secondoftwo}%
\providecommand \translation [1]{[#1]}%
\providecommand \BibitemOpen [0]{}%
\providecommand \bibitemStop [0]{}%
\providecommand \bibitemNoStop [0]{.\EOS\space}%
\providecommand \EOS [0]{\spacefactor3000\relax}%
\providecommand \BibitemShut  [1]{\csname bibitem#1\endcsname}%
\let\auto@bib@innerbib\@empty
\bibitem [{\citenamefont {Goto}\ \emph {et~al.}(2018)\citenamefont {Goto}, \citenamefont {Lin},\ and\ \citenamefont {Nakamura}}]{goto2018boltzmann}%
  \BibitemOpen
  \bibfield  {author} {\bibinfo {author} {\bibfnamefont {H.}~\bibnamefont {Goto}}, \bibinfo {author} {\bibfnamefont {Z.}~\bibnamefont {Lin}},\ and\ \bibinfo {author} {\bibfnamefont {Y.}~\bibnamefont {Nakamura}},\ }\bibfield  {title} {\emph {\bibinfo {title} {{Boltzmann sampling from the Ising model using quantum heating of coupled nonlinear oscillators}}},\ }\href@noop {} {\bibfield  {journal} {\bibinfo  {journal} {Scientific reports}\ }\textbf {\bibinfo {volume} {8}},\ \bibinfo {pages} {7154} (\bibinfo {year} {2018})}\BibitemShut {NoStop}%
\bibitem [{\citenamefont {Ponty}(2008)}]{ponty2008efficient}%
  \BibitemOpen
  \bibfield  {author} {\bibinfo {author} {\bibfnamefont {Y.}~\bibnamefont {Ponty}},\ }\bibfield  {title} {\emph {\bibinfo {title} {{Efficient sampling of RNA secondary structures from the Boltzmann ensemble of low-energy: the boustrophedon method}}},\ }\href@noop {} {\bibfield  {journal} {\bibinfo  {journal} {Journal of mathematical biology}\ }\textbf {\bibinfo {volume} {56}},\ \bibinfo {pages} {107--127} (\bibinfo {year} {2008})}\BibitemShut {NoStop}%
\bibitem [{\citenamefont {Flajolet}\ \emph {et~al.}(2007)\citenamefont {Flajolet}, \citenamefont {Fusy},\ and\ \citenamefont {Pivoteau}}]{flajolet2007boltzmann}%
  \BibitemOpen
  \bibfield  {author} {\bibinfo {author} {\bibfnamefont {P.}~\bibnamefont {Flajolet}}, \bibinfo {author} {\bibfnamefont {{\'E}.}~\bibnamefont {Fusy}},\ and\ \bibinfo {author} {\bibfnamefont {C.}~\bibnamefont {Pivoteau}},\ }in\ \href@noop {} {\emph {\bibinfo {booktitle} {2007 Proceedings of the Fourth Workshop on Analytic Algorithmics and Combinatorics (ANALCO)}}}\ (\bibinfo {organization} {SIAM},\ \bibinfo {year} {2007})\ pp.\ \bibinfo {pages} {201--211}\BibitemShut {NoStop}%
\bibitem [{\citenamefont {Augustin}\ \emph {et~al.}(1996)\citenamefont {Augustin}, \citenamefont {Mugglestone},\ and\ \citenamefont {Buckland}}]{augustin1996autologistic}%
  \BibitemOpen
  \bibfield  {author} {\bibinfo {author} {\bibfnamefont {N.}~\bibnamefont {Augustin}}, \bibinfo {author} {\bibfnamefont {M.~A.}\ \bibnamefont {Mugglestone}},\ and\ \bibinfo {author} {\bibfnamefont {S.~T.}\ \bibnamefont {Buckland}},\ }\bibfield  {title} {\emph {\bibinfo {title} {An autologistic model for the spatial distribution of wildlife}},\ }\href@noop {} {\bibfield  {journal} {\bibinfo  {journal} {Journal of Applied Ecology}\ ,\ \bibinfo {pages} {339--347}} (\bibinfo {year} {1996})}\BibitemShut {NoStop}%
\bibitem [{\citenamefont {Rohl}\ \emph {et~al.}(2004)\citenamefont {Rohl}, \citenamefont {Strauss}, \citenamefont {Misura},\ and\ \citenamefont {Baker}}]{rohl2004protein}%
  \BibitemOpen
  \bibfield  {author} {\bibinfo {author} {\bibfnamefont {C.~A.}\ \bibnamefont {Rohl}}, \bibinfo {author} {\bibfnamefont {C.~E.}\ \bibnamefont {Strauss}}, \bibinfo {author} {\bibfnamefont {K.~M.}\ \bibnamefont {Misura}},\ and\ \bibinfo {author} {\bibfnamefont {D.}~\bibnamefont {Baker}},\ }in\ \href@noop {} {\emph {\bibinfo {booktitle} {Methods in enzymology}}},\ Vol.\ \bibinfo {volume} {383}\ (\bibinfo  {publisher} {Elsevier},\ \bibinfo {year} {2004})\ pp.\ \bibinfo {pages} {66--93}\BibitemShut {NoStop}%
\bibitem [{\citenamefont {Farhi}\ \emph {et~al.}(2014)\citenamefont {Farhi}, \citenamefont {Goldstone},\ and\ \citenamefont {Gutmann}}]{QAOA}%
  \BibitemOpen
  \bibfield  {author} {\bibinfo {author} {\bibfnamefont {E.}~\bibnamefont {Farhi}}, \bibinfo {author} {\bibfnamefont {J.}~\bibnamefont {Goldstone}},\ and\ \bibinfo {author} {\bibfnamefont {S.}~\bibnamefont {Gutmann}},\ }\bibfield  {title} {\emph {\bibinfo {title} {{A Quantum Approximate Optimization Algorithm}}},\ }\href@noop {} {\bibfield  {journal} {\bibinfo  {journal} {arXiv preprint}\ } (\bibinfo {year} {2014})},\ \Eprint {https://arxiv.org/abs/1411.4028} {arXiv:1411.4028} \BibitemShut {NoStop}%
\bibitem [{\citenamefont {Farhi}\ \emph {et~al.}(2015)\citenamefont {Farhi}, \citenamefont {Goldstone},\ and\ \citenamefont {Gutmann}}]{farhi2015quantum}%
  \BibitemOpen
  \bibfield  {author} {\bibinfo {author} {\bibfnamefont {E.}~\bibnamefont {Farhi}}, \bibinfo {author} {\bibfnamefont {J.}~\bibnamefont {Goldstone}},\ and\ \bibinfo {author} {\bibfnamefont {S.}~\bibnamefont {Gutmann}},\ }\bibfield  {title} {\emph {\bibinfo {title} {{A Quantum Approximate Optimization Algorithm Applied to a Bounded Occurrence Constraint Problem}}},\ }\href@noop {} {\bibfield  {journal} {\bibinfo  {journal} {arXiv preprint}\ } (\bibinfo {year} {2015})},\ \Eprint {https://arxiv.org/abs/1412.6062} {arXiv:1412.6062} \BibitemShut {NoStop}%
\bibitem [{\citenamefont {Hadfield}\ \emph {et~al.}(2019)\citenamefont {Hadfield}, \citenamefont {Wang}, \citenamefont {O{\textquotesingle}Gorman}, \citenamefont {Rieffel}, \citenamefont {Venturelli},\ and\ \citenamefont {Biswas}}]{Hadfield_2019}%
  \BibitemOpen
  \bibfield  {author} {\bibinfo {author} {\bibfnamefont {S.}~\bibnamefont {Hadfield}}, \bibinfo {author} {\bibfnamefont {Z.}~\bibnamefont {Wang}}, \bibinfo {author} {\bibfnamefont {B.}~\bibnamefont {O{\textquotesingle}Gorman}}, \bibinfo {author} {\bibfnamefont {E.}~\bibnamefont {Rieffel}}, \bibinfo {author} {\bibfnamefont {D.}~\bibnamefont {Venturelli}},\ and\ \bibinfo {author} {\bibfnamefont {R.}~\bibnamefont {Biswas}},\ }\bibfield  {title} {\emph {\bibinfo {title} {{From the Quantum Approximate Optimization Algorithm to a Quantum Alternating Operator Ansatz}}},\ }\href {https://doi.org/10.3390/a12020034} {\bibfield  {journal} {\bibinfo  {journal} {Algorithms}\ }\textbf {\bibinfo {volume} {12}},\ \bibinfo {pages} {34} (\bibinfo {year} {2019})},\ \Eprint {https://arxiv.org/abs/1709.03489} {arXiv:1709.03489} \BibitemShut {NoStop}%
\bibitem [{\citenamefont {Wang}\ \emph {et~al.}(2020)\citenamefont {Wang}, \citenamefont {Rubin}, \citenamefont {Dominy},\ and\ \citenamefont {Rieffel}}]{PhysRevA.101.012320}%
  \BibitemOpen
  \bibfield  {author} {\bibinfo {author} {\bibfnamefont {Z.}~\bibnamefont {Wang}}, \bibinfo {author} {\bibfnamefont {N.~C.}\ \bibnamefont {Rubin}}, \bibinfo {author} {\bibfnamefont {J.~M.}\ \bibnamefont {Dominy}},\ and\ \bibinfo {author} {\bibfnamefont {E.~G.}\ \bibnamefont {Rieffel}},\ }\bibfield  {title} {\emph {\bibinfo {title} {{$XY$ mixers: Analytical and numerical results for the quantum alternating operator ansatz}}},\ }\href {https://doi.org/10.1103/PhysRevA.101.012320} {\bibfield  {journal} {\bibinfo  {journal} {Phys. Rev. A}\ }\textbf {\bibinfo {volume} {101}},\ \bibinfo {pages} {012320} (\bibinfo {year} {2020})}\BibitemShut {NoStop}%
\bibitem [{\citenamefont {Bartschi}\ and\ \citenamefont {Eidenbenz}(2020)}]{Bartschi_2020}%
  \BibitemOpen
  \bibfield  {author} {\bibinfo {author} {\bibfnamefont {A.}~\bibnamefont {Bartschi}}\ and\ \bibinfo {author} {\bibfnamefont {S.}~\bibnamefont {Eidenbenz}},\ }in\ \href {https://doi.org/10.1109/qce49297.2020.00020} {\emph {\bibinfo {booktitle} {2020 IEEE International Conference on Quantum Computing and Engineering (QCE)}}}\ (\bibinfo  {publisher} {IEEE},\ \bibinfo {year} {2020})\ p.\ \bibinfo {pages} {72–82}\BibitemShut {NoStop}%
\bibitem [{\citenamefont {He}\ \emph {et~al.}(2025)\citenamefont {He}, \citenamefont {Amaro}, \citenamefont {Shaydulin},\ and\ \citenamefont {Pistoia}}]{He_2025}%
  \BibitemOpen
  \bibfield  {author} {\bibinfo {author} {\bibfnamefont {Z.}~\bibnamefont {He}}, \bibinfo {author} {\bibfnamefont {D.}~\bibnamefont {Amaro}}, \bibinfo {author} {\bibfnamefont {R.}~\bibnamefont {Shaydulin}},\ and\ \bibinfo {author} {\bibfnamefont {M.}~\bibnamefont {Pistoia}},\ }\bibfield  {title} {\emph {\bibinfo {title} {Performance of quantum approximate optimization with quantum error detection}},\ }\bibfield  {journal} {\bibinfo  {journal} {Communications Physics}\ }\textbf {\bibinfo {volume} {8}},\ \href {https://doi.org/10.1038/s42005-025-02136-8} {10.1038/s42005-025-02136-8} (\bibinfo {year} {2025})\BibitemShut {NoStop}%
\bibitem [{\citenamefont {Harrigan}\ \emph {et~al.}(2021)\citenamefont {Harrigan}, \citenamefont {Sung}, \citenamefont {Neeley}, \citenamefont {Satzinger}, \citenamefont {Arute}, \citenamefont {Arya}, \citenamefont {Atalaya}, \citenamefont {Bardin}, \citenamefont {Barends}, \citenamefont {Boixo}, \citenamefont {Broughton}, \citenamefont {Buckley}, \citenamefont {Buell}, \citenamefont {Burkett}, \citenamefont {Bushnell}, \citenamefont {Chen}, \citenamefont {Chen}, \citenamefont {Chiaro}, \citenamefont {Collins}, \citenamefont {Courtney}, \citenamefont {Demura}, \citenamefont {Dunsworth}, \citenamefont {Eppens}, \citenamefont {Fowler}, \citenamefont {Foxen}, \citenamefont {Gidney}, \citenamefont {Giustina}, \citenamefont {Graff}, \citenamefont {Habegger}, \citenamefont {Ho}, \citenamefont {Hong}, \citenamefont {Huang}, \citenamefont {Ioffe}, \citenamefont {Isakov}, \citenamefont {Jeffrey}, \citenamefont {Jiang}, \citenamefont {Jones}, \citenamefont {Kafri}, \citenamefont {Kechedzhi}, \citenamefont {Kelly},
  \citenamefont {Kim}, \citenamefont {Klimov}, \citenamefont {Korotkov}, \citenamefont {Kostritsa}, \citenamefont {Landhuis}, \citenamefont {Laptev}, \citenamefont {Lindmark}, \citenamefont {Leib}, \citenamefont {Martin}, \citenamefont {Martinis}, \citenamefont {McClean}, \citenamefont {McEwen}, \citenamefont {Megrant}, \citenamefont {Mi}, \citenamefont {Mohseni}, \citenamefont {Mruczkiewicz}, \citenamefont {Mutus}, \citenamefont {Naaman}, \citenamefont {Neill}, \citenamefont {Neukart}, \citenamefont {Niu}, \citenamefont {O’Brien}, \citenamefont {O’Gorman}, \citenamefont {Ostby}, \citenamefont {Petukhov}, \citenamefont {Putterman}, \citenamefont {Quintana}, \citenamefont {Roushan}, \citenamefont {Rubin}, \citenamefont {Sank}, \citenamefont {Skolik}, \citenamefont {Smelyanskiy}, \citenamefont {Strain}, \citenamefont {Streif}, \citenamefont {Szalay}, \citenamefont {Vainsencher}, \citenamefont {White}, \citenamefont {Yao}, \citenamefont {Yeh}, \citenamefont {Zalcman}, \citenamefont {Zhou}, \citenamefont
  {Neven}, \citenamefont {Bacon}, \citenamefont {Lucero}, \citenamefont {Farhi},\ and\ \citenamefont {Babbush}}]{Harrigan_2021}%
  \BibitemOpen
  \bibfield  {author} {\bibinfo {author} {\bibfnamefont {M.~P.}\ \bibnamefont {Harrigan}}, \bibinfo {author} {\bibfnamefont {K.~J.}\ \bibnamefont {Sung}}, \bibinfo {author} {\bibfnamefont {M.}~\bibnamefont {Neeley}}, \bibinfo {author} {\bibfnamefont {K.~J.}\ \bibnamefont {Satzinger}}, \bibinfo {author} {\bibfnamefont {F.}~\bibnamefont {Arute}}, \bibinfo {author} {\bibfnamefont {K.}~\bibnamefont {Arya}}, \bibinfo {author} {\bibfnamefont {J.}~\bibnamefont {Atalaya}}, \bibinfo {author} {\bibfnamefont {J.~C.}\ \bibnamefont {Bardin}}, \bibinfo {author} {\bibfnamefont {R.}~\bibnamefont {Barends}}, \bibinfo {author} {\bibfnamefont {S.}~\bibnamefont {Boixo}}, \bibinfo {author} {\bibfnamefont {M.}~\bibnamefont {Broughton}}, \bibinfo {author} {\bibfnamefont {B.~B.}\ \bibnamefont {Buckley}}, \bibinfo {author} {\bibfnamefont {D.~A.}\ \bibnamefont {Buell}}, \bibinfo {author} {\bibfnamefont {B.}~\bibnamefont {Burkett}}, \bibinfo {author} {\bibfnamefont {N.}~\bibnamefont {Bushnell}}, \bibinfo {author} {\bibfnamefont
  {Y.}~\bibnamefont {Chen}}, \bibinfo {author} {\bibfnamefont {Z.}~\bibnamefont {Chen}}, \bibinfo {author} {\bibfnamefont {B.}~\bibnamefont {Chiaro}}, \bibinfo {author} {\bibfnamefont {R.}~\bibnamefont {Collins}}, \bibinfo {author} {\bibfnamefont {W.}~\bibnamefont {Courtney}}, \bibinfo {author} {\bibfnamefont {S.}~\bibnamefont {Demura}}, \bibinfo {author} {\bibfnamefont {A.}~\bibnamefont {Dunsworth}}, \bibinfo {author} {\bibfnamefont {D.}~\bibnamefont {Eppens}}, \bibinfo {author} {\bibfnamefont {A.}~\bibnamefont {Fowler}}, \bibinfo {author} {\bibfnamefont {B.}~\bibnamefont {Foxen}}, \bibinfo {author} {\bibfnamefont {C.}~\bibnamefont {Gidney}}, \bibinfo {author} {\bibfnamefont {M.}~\bibnamefont {Giustina}}, \bibinfo {author} {\bibfnamefont {R.}~\bibnamefont {Graff}}, \bibinfo {author} {\bibfnamefont {S.}~\bibnamefont {Habegger}}, \bibinfo {author} {\bibfnamefont {A.}~\bibnamefont {Ho}}, \bibinfo {author} {\bibfnamefont {S.}~\bibnamefont {Hong}}, \bibinfo {author} {\bibfnamefont {T.}~\bibnamefont {Huang}},
  \bibinfo {author} {\bibfnamefont {L.~B.}\ \bibnamefont {Ioffe}}, \bibinfo {author} {\bibfnamefont {S.~V.}\ \bibnamefont {Isakov}}, \bibinfo {author} {\bibfnamefont {E.}~\bibnamefont {Jeffrey}}, \bibinfo {author} {\bibfnamefont {Z.}~\bibnamefont {Jiang}}, \bibinfo {author} {\bibfnamefont {C.}~\bibnamefont {Jones}}, \bibinfo {author} {\bibfnamefont {D.}~\bibnamefont {Kafri}}, \bibinfo {author} {\bibfnamefont {K.}~\bibnamefont {Kechedzhi}}, \bibinfo {author} {\bibfnamefont {J.}~\bibnamefont {Kelly}}, \bibinfo {author} {\bibfnamefont {S.}~\bibnamefont {Kim}}, \bibinfo {author} {\bibfnamefont {P.~V.}\ \bibnamefont {Klimov}}, \bibinfo {author} {\bibfnamefont {A.~N.}\ \bibnamefont {Korotkov}}, \bibinfo {author} {\bibfnamefont {F.}~\bibnamefont {Kostritsa}}, \bibinfo {author} {\bibfnamefont {D.}~\bibnamefont {Landhuis}}, \bibinfo {author} {\bibfnamefont {P.}~\bibnamefont {Laptev}}, \bibinfo {author} {\bibfnamefont {M.}~\bibnamefont {Lindmark}}, \bibinfo {author} {\bibfnamefont {M.}~\bibnamefont {Leib}}, \bibinfo
  {author} {\bibfnamefont {O.}~\bibnamefont {Martin}}, \bibinfo {author} {\bibfnamefont {J.~M.}\ \bibnamefont {Martinis}}, \bibinfo {author} {\bibfnamefont {J.~R.}\ \bibnamefont {McClean}}, \bibinfo {author} {\bibfnamefont {M.}~\bibnamefont {McEwen}}, \bibinfo {author} {\bibfnamefont {A.}~\bibnamefont {Megrant}}, \bibinfo {author} {\bibfnamefont {X.}~\bibnamefont {Mi}}, \bibinfo {author} {\bibfnamefont {M.}~\bibnamefont {Mohseni}}, \bibinfo {author} {\bibfnamefont {W.}~\bibnamefont {Mruczkiewicz}}, \bibinfo {author} {\bibfnamefont {J.}~\bibnamefont {Mutus}}, \bibinfo {author} {\bibfnamefont {O.}~\bibnamefont {Naaman}}, \bibinfo {author} {\bibfnamefont {C.}~\bibnamefont {Neill}}, \bibinfo {author} {\bibfnamefont {F.}~\bibnamefont {Neukart}}, \bibinfo {author} {\bibfnamefont {M.~Y.}\ \bibnamefont {Niu}}, \bibinfo {author} {\bibfnamefont {T.~E.}\ \bibnamefont {O’Brien}}, \bibinfo {author} {\bibfnamefont {B.}~\bibnamefont {O’Gorman}}, \bibinfo {author} {\bibfnamefont {E.}~\bibnamefont {Ostby}}, \bibinfo
  {author} {\bibfnamefont {A.}~\bibnamefont {Petukhov}}, \bibinfo {author} {\bibfnamefont {H.}~\bibnamefont {Putterman}}, \bibinfo {author} {\bibfnamefont {C.}~\bibnamefont {Quintana}}, \bibinfo {author} {\bibfnamefont {P.}~\bibnamefont {Roushan}}, \bibinfo {author} {\bibfnamefont {N.~C.}\ \bibnamefont {Rubin}}, \bibinfo {author} {\bibfnamefont {D.}~\bibnamefont {Sank}}, \bibinfo {author} {\bibfnamefont {A.}~\bibnamefont {Skolik}}, \bibinfo {author} {\bibfnamefont {V.}~\bibnamefont {Smelyanskiy}}, \bibinfo {author} {\bibfnamefont {D.}~\bibnamefont {Strain}}, \bibinfo {author} {\bibfnamefont {M.}~\bibnamefont {Streif}}, \bibinfo {author} {\bibfnamefont {M.}~\bibnamefont {Szalay}}, \bibinfo {author} {\bibfnamefont {A.}~\bibnamefont {Vainsencher}}, \bibinfo {author} {\bibfnamefont {T.}~\bibnamefont {White}}, \bibinfo {author} {\bibfnamefont {Z.~J.}\ \bibnamefont {Yao}}, \bibinfo {author} {\bibfnamefont {P.}~\bibnamefont {Yeh}}, \bibinfo {author} {\bibfnamefont {A.}~\bibnamefont {Zalcman}}, \bibinfo {author}
  {\bibfnamefont {L.}~\bibnamefont {Zhou}}, \bibinfo {author} {\bibfnamefont {H.}~\bibnamefont {Neven}}, \bibinfo {author} {\bibfnamefont {D.}~\bibnamefont {Bacon}}, \bibinfo {author} {\bibfnamefont {E.}~\bibnamefont {Lucero}}, \bibinfo {author} {\bibfnamefont {E.}~\bibnamefont {Farhi}},\ and\ \bibinfo {author} {\bibfnamefont {R.}~\bibnamefont {Babbush}},\ }\bibfield  {title} {\emph {\bibinfo {title} {Quantum approximate optimization of non-planar graph problems on a planar superconducting processor}},\ }\href {https://doi.org/10.1038/s41567-020-01105-y} {\bibfield  {journal} {\bibinfo  {journal} {Nature Physics}\ }\textbf {\bibinfo {volume} {17}},\ \bibinfo {pages} {332–336} (\bibinfo {year} {2021})}\BibitemShut {NoStop}%
\bibitem [{\citenamefont {Pelofske}\ \emph {et~al.}(2024{\natexlab{a}})\citenamefont {Pelofske}, \citenamefont {Bärtschi}, \citenamefont {Cincio}, \citenamefont {Golden},\ and\ \citenamefont {Eidenbenz}}]{Pelofske_2024_scaling_QAOA}%
  \BibitemOpen
  \bibfield  {author} {\bibinfo {author} {\bibfnamefont {E.}~\bibnamefont {Pelofske}}, \bibinfo {author} {\bibfnamefont {A.}~\bibnamefont {Bärtschi}}, \bibinfo {author} {\bibfnamefont {L.}~\bibnamefont {Cincio}}, \bibinfo {author} {\bibfnamefont {J.}~\bibnamefont {Golden}},\ and\ \bibinfo {author} {\bibfnamefont {S.}~\bibnamefont {Eidenbenz}},\ }\bibfield  {title} {\emph {\bibinfo {title} {{Scaling whole-chip QAOA for higher-order ising spin glass models on heavy-hex graphs}}},\ }\bibfield  {journal} {\bibinfo  {journal} {npj Quantum Information}\ }\textbf {\bibinfo {volume} {10}},\ \href {https://doi.org/10.1038/s41534-024-00906-w} {10.1038/s41534-024-00906-w} (\bibinfo {year} {2024}{\natexlab{a}})\BibitemShut {NoStop}%
\bibitem [{\citenamefont {Pelofske}\ \emph {et~al.}(2024{\natexlab{b}})\citenamefont {Pelofske}, \citenamefont {B{\"a}rtschi},\ and\ \citenamefont {Eidenbenz}}]{pelofske2023short}%
  \BibitemOpen
  \bibfield  {author} {\bibinfo {author} {\bibfnamefont {E.}~\bibnamefont {Pelofske}}, \bibinfo {author} {\bibfnamefont {A.}~\bibnamefont {B{\"a}rtschi}},\ and\ \bibinfo {author} {\bibfnamefont {S.}~\bibnamefont {Eidenbenz}},\ }\bibfield  {title} {\emph {\bibinfo {title} {{Short-Depth QAOA Circuits and Quantum Annealing on Higher-Order Ising Models}}},\ }\href {https://doi.org/10.1038/s41534-024-00825-w} {\bibfield  {journal} {\bibinfo  {journal} {npj Quantum Information}\ }\textbf {\bibinfo {volume} {10}},\ \bibinfo {pages} {30} (\bibinfo {year} {2024}{\natexlab{b}})}\BibitemShut {NoStop}%
\bibitem [{\citenamefont {Shaydulin}\ and\ \citenamefont {Pistoia}(2023)}]{shaydulin2023qaoancdotpgeq200}%
  \BibitemOpen
  \bibfield  {author} {\bibinfo {author} {\bibfnamefont {R.}~\bibnamefont {Shaydulin}}\ and\ \bibinfo {author} {\bibfnamefont {M.}~\bibnamefont {Pistoia}},\ }\href {https://arxiv.org/abs/2303.02064} {\bibinfo {title} {{QAOA with $N\cdot p\geq 200$}}} (\bibinfo {year} {2023}),\ \Eprint {https://arxiv.org/abs/2303.02064} {arXiv:2303.02064 [quant-ph]} \BibitemShut {NoStop}%
\bibitem [{\citenamefont {Pelofske}\ \emph {et~al.}(2023)\citenamefont {Pelofske}, \citenamefont {Bärtschi},\ and\ \citenamefont {Eidenbenz}}]{Pelofske_2023}%
  \BibitemOpen
  \bibfield  {author} {\bibinfo {author} {\bibfnamefont {E.}~\bibnamefont {Pelofske}}, \bibinfo {author} {\bibfnamefont {A.}~\bibnamefont {Bärtschi}},\ and\ \bibinfo {author} {\bibfnamefont {S.}~\bibnamefont {Eidenbenz}},\ }\bibinfo {title} {{Quantum Annealing vs. QAOA: 127 Qubit Higher-Order Ising Problems on NISQ Computers}},\ in\ \href {https://doi.org/10.1007/978-3-031-32041-5_13} {\emph {\bibinfo {booktitle} {High Performance Computing}}}\ (\bibinfo  {publisher} {Springer Nature Switzerland},\ \bibinfo {year} {2023})\ p.\ \bibinfo {pages} {240–258}\BibitemShut {NoStop}%
\bibitem [{\citenamefont {Weidenfeller}\ \emph {et~al.}(2022)\citenamefont {Weidenfeller}, \citenamefont {Valor}, \citenamefont {Gacon}, \citenamefont {Tornow}, \citenamefont {Bello}, \citenamefont {Woerner},\ and\ \citenamefont {Egger}}]{Weidenfeller_2022}%
  \BibitemOpen
  \bibfield  {author} {\bibinfo {author} {\bibfnamefont {J.}~\bibnamefont {Weidenfeller}}, \bibinfo {author} {\bibfnamefont {L.~C.}\ \bibnamefont {Valor}}, \bibinfo {author} {\bibfnamefont {J.}~\bibnamefont {Gacon}}, \bibinfo {author} {\bibfnamefont {C.}~\bibnamefont {Tornow}}, \bibinfo {author} {\bibfnamefont {L.}~\bibnamefont {Bello}}, \bibinfo {author} {\bibfnamefont {S.}~\bibnamefont {Woerner}},\ and\ \bibinfo {author} {\bibfnamefont {D.~J.}\ \bibnamefont {Egger}},\ }\bibfield  {title} {\emph {\bibinfo {title} {Scaling of the quantum approximate optimization algorithm on superconducting qubit based hardware}},\ }\href {https://doi.org/10.22331/q-2022-12-07-870} {\bibfield  {journal} {\bibinfo  {journal} {Quantum}\ }\textbf {\bibinfo {volume} {6}},\ \bibinfo {pages} {870} (\bibinfo {year} {2022})}\BibitemShut {NoStop}%
\bibitem [{\citenamefont {Golden}\ \emph {et~al.}(2022)\citenamefont {Golden}, \citenamefont {Bärtschi}, \citenamefont {O’Malley},\ and\ \citenamefont {Eidenbenz}}]{Golden_2022_fair}%
  \BibitemOpen
  \bibfield  {author} {\bibinfo {author} {\bibfnamefont {J.}~\bibnamefont {Golden}}, \bibinfo {author} {\bibfnamefont {A.}~\bibnamefont {Bärtschi}}, \bibinfo {author} {\bibfnamefont {D.}~\bibnamefont {O’Malley}},\ and\ \bibinfo {author} {\bibfnamefont {S.}~\bibnamefont {Eidenbenz}},\ }\bibfield  {title} {\emph {\bibinfo {title} {{Fair Sampling Error Analysis on NISQ Devices}}},\ }\href {https://doi.org/10.1145/3510857} {\bibfield  {journal} {\bibinfo  {journal} {ACM Transactions on Quantum Computing}\ }\textbf {\bibinfo {volume} {3}},\ \bibinfo {pages} {1–23} (\bibinfo {year} {2022})}\BibitemShut {NoStop}%
\bibitem [{\citenamefont {Pelofske}\ \emph {et~al.}(2021)\citenamefont {Pelofske}, \citenamefont {Golden}, \citenamefont {Bartschi}, \citenamefont {O’Malley},\ and\ \citenamefont {Eidenbenz}}]{Pelofske_2021}%
  \BibitemOpen
  \bibfield  {author} {\bibinfo {author} {\bibfnamefont {E.}~\bibnamefont {Pelofske}}, \bibinfo {author} {\bibfnamefont {J.}~\bibnamefont {Golden}}, \bibinfo {author} {\bibfnamefont {A.}~\bibnamefont {Bartschi}}, \bibinfo {author} {\bibfnamefont {D.}~\bibnamefont {O’Malley}},\ and\ \bibinfo {author} {\bibfnamefont {S.}~\bibnamefont {Eidenbenz}},\ }in\ \href {https://doi.org/10.1109/qce52317.2021.00038} {\emph {\bibinfo {booktitle} {2021 IEEE International Conference on Quantum Computing and Engineering (QCE)}}}\ (\bibinfo  {publisher} {IEEE},\ \bibinfo {year} {2021})\ p.\ \bibinfo {pages} {207–217}\BibitemShut {NoStop}%
\bibitem [{\citenamefont {Pelofske}(2025{\natexlab{a}})}]{Pelofske_2025}%
  \BibitemOpen
  \bibfield  {author} {\bibinfo {author} {\bibfnamefont {E.}~\bibnamefont {Pelofske}},\ }\bibfield  {title} {\emph {\bibinfo {title} {{Biased degenerate ground-state sampling of small Ising models with converged quantum approximate optimization algorithm}}},\ }\bibfield  {journal} {\bibinfo  {journal} {Physical Review E}\ }\textbf {\bibinfo {volume} {111}},\ \href {https://doi.org/10.1103/physreve.111.054103} {10.1103/physreve.111.054103} (\bibinfo {year} {2025}{\natexlab{a}})\BibitemShut {NoStop}%
\bibitem [{\citenamefont {Albash}\ \emph {et~al.}(2015)\citenamefont {Albash}, \citenamefont {Vinci}, \citenamefont {Mishra}, \citenamefont {Warburton},\ and\ \citenamefont {Lidar}}]{Albash_2015}%
  \BibitemOpen
  \bibfield  {author} {\bibinfo {author} {\bibfnamefont {T.}~\bibnamefont {Albash}}, \bibinfo {author} {\bibfnamefont {W.}~\bibnamefont {Vinci}}, \bibinfo {author} {\bibfnamefont {A.}~\bibnamefont {Mishra}}, \bibinfo {author} {\bibfnamefont {P.~A.}\ \bibnamefont {Warburton}},\ and\ \bibinfo {author} {\bibfnamefont {D.~A.}\ \bibnamefont {Lidar}},\ }\bibfield  {title} {\emph {\bibinfo {title} {Consistency tests of classical and quantum models for a quantum annealer}},\ }\bibfield  {journal} {\bibinfo  {journal} {Physical Review A}\ }\textbf {\bibinfo {volume} {91}},\ \href {https://doi.org/10.1103/physreva.91.042314} {10.1103/physreva.91.042314} (\bibinfo {year} {2015})\BibitemShut {NoStop}%
\bibitem [{\citenamefont {Matsuda}\ \emph {et~al.}(2009)\citenamefont {Matsuda}, \citenamefont {Nishimori},\ and\ \citenamefont {Katzgraber}}]{matsuda2009quantum}%
  \BibitemOpen
  \bibfield  {author} {\bibinfo {author} {\bibfnamefont {Y.}~\bibnamefont {Matsuda}}, \bibinfo {author} {\bibfnamefont {H.}~\bibnamefont {Nishimori}},\ and\ \bibinfo {author} {\bibfnamefont {H.~G.}\ \bibnamefont {Katzgraber}},\ }in\ \href {https://doi.org/10.1088/1742-6596/143/1/012003} {\emph {\bibinfo {booktitle} {Journal of Physics: Conference Series}}},\ Vol.\ \bibinfo {volume} {143}\ (\bibinfo {organization} {IOP Publishing},\ \bibinfo {year} {2009})\ p.\ \bibinfo {pages} {012003}\BibitemShut {NoStop}%
\bibitem [{\citenamefont {K\"onz}\ \emph {et~al.}(2019)\citenamefont {K\"onz}, \citenamefont {Mazzola}, \citenamefont {Ochoa}, \citenamefont {Katzgraber},\ and\ \citenamefont {Troyer}}]{PhysRevA.100.030303}%
  \BibitemOpen
  \bibfield  {author} {\bibinfo {author} {\bibfnamefont {M.~S.}\ \bibnamefont {K\"onz}}, \bibinfo {author} {\bibfnamefont {G.}~\bibnamefont {Mazzola}}, \bibinfo {author} {\bibfnamefont {A.~J.}\ \bibnamefont {Ochoa}}, \bibinfo {author} {\bibfnamefont {H.~G.}\ \bibnamefont {Katzgraber}},\ and\ \bibinfo {author} {\bibfnamefont {M.}~\bibnamefont {Troyer}},\ }\bibfield  {title} {\emph {\bibinfo {title} {Uncertain fate of fair sampling in quantum annealing}},\ }\href {https://doi.org/10.1103/PhysRevA.100.030303} {\bibfield  {journal} {\bibinfo  {journal} {Phys. Rev. A}\ }\textbf {\bibinfo {volume} {100}},\ \bibinfo {pages} {030303} (\bibinfo {year} {2019})}\BibitemShut {NoStop}%
\bibitem [{\citenamefont {Mandr\`a}\ \emph {et~al.}(2017)\citenamefont {Mandr\`a}, \citenamefont {Zhu},\ and\ \citenamefont {Katzgraber}}]{PhysRevLett.118.070502}%
  \BibitemOpen
  \bibfield  {author} {\bibinfo {author} {\bibfnamefont {S.}~\bibnamefont {Mandr\`a}}, \bibinfo {author} {\bibfnamefont {Z.}~\bibnamefont {Zhu}},\ and\ \bibinfo {author} {\bibfnamefont {H.~G.}\ \bibnamefont {Katzgraber}},\ }\bibfield  {title} {\emph {\bibinfo {title} {Exponentially biased ground-state sampling of quantum annealing machines with transverse-field driving hamiltonians}},\ }\href {https://doi.org/10.1103/PhysRevLett.118.070502} {\bibfield  {journal} {\bibinfo  {journal} {Phys. Rev. Lett.}\ }\textbf {\bibinfo {volume} {118}},\ \bibinfo {pages} {070502} (\bibinfo {year} {2017})}\BibitemShut {NoStop}%
\bibitem [{\citenamefont {Zhu}\ \emph {et~al.}(2019)\citenamefont {Zhu}, \citenamefont {Ochoa},\ and\ \citenamefont {Katzgraber}}]{PhysRevE.99.063314}%
  \BibitemOpen
  \bibfield  {author} {\bibinfo {author} {\bibfnamefont {Z.}~\bibnamefont {Zhu}}, \bibinfo {author} {\bibfnamefont {A.~J.}\ \bibnamefont {Ochoa}},\ and\ \bibinfo {author} {\bibfnamefont {H.~G.}\ \bibnamefont {Katzgraber}},\ }\bibfield  {title} {\emph {\bibinfo {title} {Fair sampling of ground-state configurations of binary optimization problems}},\ }\href {https://doi.org/10.1103/PhysRevE.99.063314} {\bibfield  {journal} {\bibinfo  {journal} {Phys. Rev. E}\ }\textbf {\bibinfo {volume} {99}},\ \bibinfo {pages} {063314} (\bibinfo {year} {2019})}\BibitemShut {NoStop}%
\bibitem [{\citenamefont {Farhi}\ and\ \citenamefont {Harrow}(2019)}]{farhi2019quantumsupremacyquantumapproximate}%
  \BibitemOpen
  \bibfield  {author} {\bibinfo {author} {\bibfnamefont {E.}~\bibnamefont {Farhi}}\ and\ \bibinfo {author} {\bibfnamefont {A.~W.}\ \bibnamefont {Harrow}},\ }\href {https://arxiv.org/abs/1602.07674} {\bibinfo {title} {Quantum supremacy through the quantum approximate optimization algorithm}} (\bibinfo {year} {2019}),\ \Eprint {https://arxiv.org/abs/1602.07674} {arXiv:1602.07674 [quant-ph]} \BibitemShut {NoStop}%
\bibitem [{\citenamefont {Krovi}(2022)}]{krovi2022averagecasehardnessestimatingprobabilities}%
  \BibitemOpen
  \bibfield  {author} {\bibinfo {author} {\bibfnamefont {H.}~\bibnamefont {Krovi}},\ }\href {https://arxiv.org/abs/2206.05642} {\bibinfo {title} {Average-case hardness of estimating probabilities of random quantum circuits with a linear scaling in the error exponent}} (\bibinfo {year} {2022}),\ \Eprint {https://arxiv.org/abs/2206.05642} {arXiv:2206.05642 [quant-ph]} \BibitemShut {NoStop}%
\bibitem [{\citenamefont {Farhi}\ \emph {et~al.}(2000)\citenamefont {Farhi}, \citenamefont {Goldstone}, \citenamefont {Gutmann},\ and\ \citenamefont {Sipser}}]{farhi2000quantumcomputationadiabaticevolution}%
  \BibitemOpen
  \bibfield  {author} {\bibinfo {author} {\bibfnamefont {E.}~\bibnamefont {Farhi}}, \bibinfo {author} {\bibfnamefont {J.}~\bibnamefont {Goldstone}}, \bibinfo {author} {\bibfnamefont {S.}~\bibnamefont {Gutmann}},\ and\ \bibinfo {author} {\bibfnamefont {M.}~\bibnamefont {Sipser}},\ }\href {https://arxiv.org/abs/quant-ph/0001106} {\bibinfo {title} {{Quantum Computation by Adiabatic Evolution}}} (\bibinfo {year} {2000}),\ \Eprint {https://arxiv.org/abs/quant-ph/0001106} {arXiv:quant-ph/0001106 [quant-ph]} \BibitemShut {NoStop}%
\bibitem [{\citenamefont {Sack}\ and\ \citenamefont {Serbyn}(2021)}]{Sack_2021}%
  \BibitemOpen
  \bibfield  {author} {\bibinfo {author} {\bibfnamefont {S.~H.}\ \bibnamefont {Sack}}\ and\ \bibinfo {author} {\bibfnamefont {M.}~\bibnamefont {Serbyn}},\ }\bibfield  {title} {\emph {\bibinfo {title} {Quantum annealing initialization of the quantum approximate optimization algorithm}},\ }\href {https://doi.org/10.22331/q-2021-07-01-491} {\bibfield  {journal} {\bibinfo  {journal} {Quantum}\ }\textbf {\bibinfo {volume} {5}},\ \bibinfo {pages} {491} (\bibinfo {year} {2021})},\ \Eprint {https://arxiv.org/abs/2101.05742} {arXiv:2101.05742} \BibitemShut {NoStop}%
\bibitem [{\citenamefont {Boulebnane}\ \emph {et~al.}(2025)\citenamefont {Boulebnane}, \citenamefont {Sud}, \citenamefont {Shaydulin},\ and\ \citenamefont {Pistoia}}]{boulebnane2025equivalencequantumapproximateoptimization}%
  \BibitemOpen
  \bibfield  {author} {\bibinfo {author} {\bibfnamefont {S.}~\bibnamefont {Boulebnane}}, \bibinfo {author} {\bibfnamefont {J.}~\bibnamefont {Sud}}, \bibinfo {author} {\bibfnamefont {R.}~\bibnamefont {Shaydulin}},\ and\ \bibinfo {author} {\bibfnamefont {M.}~\bibnamefont {Pistoia}},\ }\bibfield  {title} {\emph {\bibinfo {title} {{Equivalence of Quantum Approximate Optimization Algorithm and Linear-Time Quantum Annealing for the Sherrington-Kirkpatrick Model}}},\ }\href@noop {} {\bibfield  {journal} {\bibinfo  {journal} {arXiv preprint}\ } (\bibinfo {year} {2025})},\ \Eprint {https://arxiv.org/abs/2503.09563} {arXiv:2503.09563} \BibitemShut {NoStop}%
\bibitem [{\citenamefont {Mbeng}\ \emph {et~al.}(2019)\citenamefont {Mbeng}, \citenamefont {Fazio},\ and\ \citenamefont {Santoro}}]{mbeng2019quantumannealingjourneydigitalization}%
  \BibitemOpen
  \bibfield  {author} {\bibinfo {author} {\bibfnamefont {G.~B.}\ \bibnamefont {Mbeng}}, \bibinfo {author} {\bibfnamefont {R.}~\bibnamefont {Fazio}},\ and\ \bibinfo {author} {\bibfnamefont {G.}~\bibnamefont {Santoro}},\ }\bibfield  {title} {\emph {\bibinfo {title} {{Quantum Annealing: a journey through Digitalization, Control, and hybrid Quantum Variational schemes}}},\ }\href@noop {} {\bibfield  {journal} {\bibinfo  {journal} {arXiv preprint}\ } (\bibinfo {year} {2019})},\ \Eprint {https://arxiv.org/abs/1906.08948} {arXiv:1906.08948} \BibitemShut {NoStop}%
\bibitem [{\citenamefont {Kovalsky}\ \emph {et~al.}(2023)\citenamefont {Kovalsky}, \citenamefont {Calderon-Vargas}, \citenamefont {Grace}, \citenamefont {Magann}, \citenamefont {Larsen}, \citenamefont {Baczewski},\ and\ \citenamefont {Sarovar}}]{Kovalsky_2023}%
  \BibitemOpen
  \bibfield  {author} {\bibinfo {author} {\bibfnamefont {L.~K.}\ \bibnamefont {Kovalsky}}, \bibinfo {author} {\bibfnamefont {F.~A.}\ \bibnamefont {Calderon-Vargas}}, \bibinfo {author} {\bibfnamefont {M.~D.}\ \bibnamefont {Grace}}, \bibinfo {author} {\bibfnamefont {A.~B.}\ \bibnamefont {Magann}}, \bibinfo {author} {\bibfnamefont {J.~B.}\ \bibnamefont {Larsen}}, \bibinfo {author} {\bibfnamefont {A.~D.}\ \bibnamefont {Baczewski}},\ and\ \bibinfo {author} {\bibfnamefont {M.}~\bibnamefont {Sarovar}},\ }\bibfield  {title} {\emph {\bibinfo {title} {{Self-Healing of Trotter Error in Digital Adiabatic State Preparation}}},\ }\href {https://doi.org/10.1103/physrevlett.131.060602} {\bibfield  {journal} {\bibinfo  {journal} {Physical Review Letters}\ }\textbf {\bibinfo {volume} {131}},\ \bibinfo {pages} {060602} (\bibinfo {year} {2023})},\ \Eprint {https://arxiv.org/abs/2209.06242} {arXiv:2209.06242} \BibitemShut {NoStop}%
\bibitem [{\citenamefont {Montañez-Barrera}\ and\ \citenamefont {Michielsen}(2025)}]{Monta_ez_Barrera_2025}%
  \BibitemOpen
  \bibfield  {author} {\bibinfo {author} {\bibfnamefont {J.~A.}\ \bibnamefont {Montañez-Barrera}}\ and\ \bibinfo {author} {\bibfnamefont {K.}~\bibnamefont {Michielsen}},\ }\bibfield  {title} {\emph {\bibinfo {title} {{Toward a linear-ramp QAOA protocol: evidence of a scaling advantage in solving some combinatorial optimization problems}}},\ }\href {https://doi.org/10.1038/s41534-025-01082-1} {\bibfield  {journal} {\bibinfo  {journal} {npj Quantum Information}\ }\textbf {\bibinfo {volume} {11}},\ \bibinfo {pages} {131} (\bibinfo {year} {2025})},\ \Eprint {https://arxiv.org/abs/2405.09169} {arXiv:2405.09169} \BibitemShut {NoStop}%
\bibitem [{\citenamefont {Grover}(1996)}]{grover1996fastquantummechanicalalgorithm}%
  \BibitemOpen
  \bibfield  {author} {\bibinfo {author} {\bibfnamefont {L.~K.}\ \bibnamefont {Grover}},\ }\href {https://arxiv.org/abs/quant-ph/9605043} {\bibinfo {title} {A fast quantum mechanical algorithm for database search}} (\bibinfo {year} {1996}),\ \Eprint {https://arxiv.org/abs/quant-ph/9605043} {arXiv:quant-ph/9605043 [quant-ph]} \BibitemShut {NoStop}%
\bibitem [{\citenamefont {Ball}\ and\ \citenamefont {Cohen}(2023)}]{Ball_2023}%
  \BibitemOpen
  \bibfield  {author} {\bibinfo {author} {\bibfnamefont {C.}~\bibnamefont {Ball}}\ and\ \bibinfo {author} {\bibfnamefont {T.~D.}\ \bibnamefont {Cohen}},\ }\bibfield  {title} {\emph {\bibinfo {title} {Boltzmann distributions on a quantum computer via active cooling}},\ }\href {https://doi.org/10.1016/j.nuclphysa.2023.122708} {\bibfield  {journal} {\bibinfo  {journal} {Nuclear Physics A}\ }\textbf {\bibinfo {volume} {1038}},\ \bibinfo {pages} {122708} (\bibinfo {year} {2023})}\BibitemShut {NoStop}%
\bibitem [{\citenamefont {Marshall}\ \emph {et~al.}(2019)\citenamefont {Marshall}, \citenamefont {Venturelli}, \citenamefont {Hen},\ and\ \citenamefont {Rieffel}}]{PhysRevApplied.11.044083}%
  \BibitemOpen
  \bibfield  {author} {\bibinfo {author} {\bibfnamefont {J.}~\bibnamefont {Marshall}}, \bibinfo {author} {\bibfnamefont {D.}~\bibnamefont {Venturelli}}, \bibinfo {author} {\bibfnamefont {I.}~\bibnamefont {Hen}},\ and\ \bibinfo {author} {\bibfnamefont {E.~G.}\ \bibnamefont {Rieffel}},\ }\bibfield  {title} {\emph {\bibinfo {title} {{Power of Pausing: Advancing Understanding of Thermalization in Experimental Quantum Annealers}}},\ }\href {https://doi.org/10.1103/PhysRevApplied.11.044083} {\bibfield  {journal} {\bibinfo  {journal} {Phys. Rev. Appl.}\ }\textbf {\bibinfo {volume} {11}},\ \bibinfo {pages} {044083} (\bibinfo {year} {2019})}\BibitemShut {NoStop}%
\bibitem [{\citenamefont {Izquierdo}\ \emph {et~al.}(2021)\citenamefont {Izquierdo}, \citenamefont {Hen},\ and\ \citenamefont {Albash}}]{Izquierdo_2021}%
  \BibitemOpen
  \bibfield  {author} {\bibinfo {author} {\bibfnamefont {Z.~G.}\ \bibnamefont {Izquierdo}}, \bibinfo {author} {\bibfnamefont {I.}~\bibnamefont {Hen}},\ and\ \bibinfo {author} {\bibfnamefont {T.}~\bibnamefont {Albash}},\ }\bibfield  {title} {\emph {\bibinfo {title} {{Testing a Quantum Annealer as a Quantum Thermal Sampler}}},\ }\href {https://doi.org/10.1145/3464456} {\bibfield  {journal} {\bibinfo  {journal} {{ACM} Transactions on Quantum Computing}\ }\textbf {\bibinfo {volume} {2}},\ \bibinfo {pages} {1--20} (\bibinfo {year} {2021})},\ \Eprint {https://arxiv.org/abs/2003.00361} {arXiv:2003.00361} \BibitemShut {NoStop}%
\bibitem [{\citenamefont {Marshall}\ \emph {et~al.}(2017)\citenamefont {Marshall}, \citenamefont {Rieffel},\ and\ \citenamefont {Hen}}]{Marshall_2017}%
  \BibitemOpen
  \bibfield  {author} {\bibinfo {author} {\bibfnamefont {J.}~\bibnamefont {Marshall}}, \bibinfo {author} {\bibfnamefont {E.~G.}\ \bibnamefont {Rieffel}},\ and\ \bibinfo {author} {\bibfnamefont {I.}~\bibnamefont {Hen}},\ }\bibfield  {title} {\emph {\bibinfo {title} {{Thermalization, Freeze-out, and Noise: Deciphering Experimental Quantum Annealers}}},\ }\bibfield  {journal} {\bibinfo  {journal} {Physical Review Applied}\ }\textbf {\bibinfo {volume} {8}},\ \href {https://doi.org/10.1103/physrevapplied.8.064025} {10.1103/physrevapplied.8.064025} (\bibinfo {year} {2017})\BibitemShut {NoStop}%
\bibitem [{\citenamefont {Nelson}\ \emph {et~al.}(2021)\citenamefont {Nelson}, \citenamefont {Vuffray}, \citenamefont {Lokhov},\ and\ \citenamefont {Coffrin}}]{nelson2021singlequbitfidelityassessmentquantum}%
  \BibitemOpen
  \bibfield  {author} {\bibinfo {author} {\bibfnamefont {J.}~\bibnamefont {Nelson}}, \bibinfo {author} {\bibfnamefont {M.}~\bibnamefont {Vuffray}}, \bibinfo {author} {\bibfnamefont {A.~Y.}\ \bibnamefont {Lokhov}},\ and\ \bibinfo {author} {\bibfnamefont {C.}~\bibnamefont {Coffrin}},\ }\href {https://arxiv.org/abs/2104.03335} {\bibinfo {title} {{Single-Qubit Fidelity Assessment of Quantum Annealing Hardware}}} (\bibinfo {year} {2021}),\ \Eprint {https://arxiv.org/abs/2104.03335} {arXiv:2104.03335 [quant-ph]} \BibitemShut {NoStop}%
\bibitem [{\citenamefont {Vuffray}\ \emph {et~al.}(2022)\citenamefont {Vuffray}, \citenamefont {Coffrin}, \citenamefont {Kharkov},\ and\ \citenamefont {Lokhov}}]{PRXQuantum.3.020317}%
  \BibitemOpen
  \bibfield  {author} {\bibinfo {author} {\bibfnamefont {M.}~\bibnamefont {Vuffray}}, \bibinfo {author} {\bibfnamefont {C.}~\bibnamefont {Coffrin}}, \bibinfo {author} {\bibfnamefont {Y.~A.}\ \bibnamefont {Kharkov}},\ and\ \bibinfo {author} {\bibfnamefont {A.~Y.}\ \bibnamefont {Lokhov}},\ }\bibfield  {title} {\emph {\bibinfo {title} {{Programmable Quantum Annealers as Noisy Gibbs Samplers}}},\ }\href {https://doi.org/10.1103/PRXQuantum.3.020317} {\bibfield  {journal} {\bibinfo  {journal} {PRX Quantum}\ }\textbf {\bibinfo {volume} {3}},\ \bibinfo {pages} {020317} (\bibinfo {year} {2022})}\BibitemShut {NoStop}%
\bibitem [{\citenamefont {Nelson}\ \emph {et~al.}(2022)\citenamefont {Nelson}, \citenamefont {Vuffray}, \citenamefont {Lokhov}, \citenamefont {Albash},\ and\ \citenamefont {Coffrin}}]{PhysRevApplied.17.044046}%
  \BibitemOpen
  \bibfield  {author} {\bibinfo {author} {\bibfnamefont {J.}~\bibnamefont {Nelson}}, \bibinfo {author} {\bibfnamefont {M.}~\bibnamefont {Vuffray}}, \bibinfo {author} {\bibfnamefont {A.~Y.}\ \bibnamefont {Lokhov}}, \bibinfo {author} {\bibfnamefont {T.}~\bibnamefont {Albash}},\ and\ \bibinfo {author} {\bibfnamefont {C.}~\bibnamefont {Coffrin}},\ }\bibfield  {title} {\emph {\bibinfo {title} {{High-Quality Thermal Gibbs Sampling with Quantum Annealing Hardware}}},\ }\href {https://doi.org/10.1103/PhysRevApplied.17.044046} {\bibfield  {journal} {\bibinfo  {journal} {Phys. Rev. Appl.}\ }\textbf {\bibinfo {volume} {17}},\ \bibinfo {pages} {044046} (\bibinfo {year} {2022})}\BibitemShut {NoStop}%
\bibitem [{\citenamefont {Buffoni}\ and\ \citenamefont {Campisi}(2020)}]{buffoni2020thermodynamics}%
  \BibitemOpen
  \bibfield  {author} {\bibinfo {author} {\bibfnamefont {L.}~\bibnamefont {Buffoni}}\ and\ \bibinfo {author} {\bibfnamefont {M.}~\bibnamefont {Campisi}},\ }\bibfield  {title} {\emph {\bibinfo {title} {Thermodynamics of a quantum annealer}},\ }\href@noop {} {\bibfield  {journal} {\bibinfo  {journal} {Quantum Science and Technology}\ }\textbf {\bibinfo {volume} {5}},\ \bibinfo {pages} {035013} (\bibinfo {year} {2020})}\BibitemShut {NoStop}%
\bibitem [{\citenamefont {Mörstedt}\ \emph {et~al.}(2024)\citenamefont {Mörstedt}, \citenamefont {Teixeira}, \citenamefont {Viitanen}, \citenamefont {Kivijärvi}, \citenamefont {Tiiri}, \citenamefont {Rasola}, \citenamefont {Gunyho}, \citenamefont {Kundu}, \citenamefont {Lattier}, \citenamefont {Vadimov}, \citenamefont {Catelani}, \citenamefont {Sevriuk}, \citenamefont {Heinsoo}, \citenamefont {Räbinä}, \citenamefont {Ankerhold},\ and\ \citenamefont {Möttönen}}]{mörstedt2024rapidondemandgenerationthermal}%
  \BibitemOpen
  \bibfield  {author} {\bibinfo {author} {\bibfnamefont {T.~F.}\ \bibnamefont {Mörstedt}}, \bibinfo {author} {\bibfnamefont {W.~S.}\ \bibnamefont {Teixeira}}, \bibinfo {author} {\bibfnamefont {A.}~\bibnamefont {Viitanen}}, \bibinfo {author} {\bibfnamefont {H.}~\bibnamefont {Kivijärvi}}, \bibinfo {author} {\bibfnamefont {M.}~\bibnamefont {Tiiri}}, \bibinfo {author} {\bibfnamefont {M.}~\bibnamefont {Rasola}}, \bibinfo {author} {\bibfnamefont {A.~M.}\ \bibnamefont {Gunyho}}, \bibinfo {author} {\bibfnamefont {S.}~\bibnamefont {Kundu}}, \bibinfo {author} {\bibfnamefont {L.}~\bibnamefont {Lattier}}, \bibinfo {author} {\bibfnamefont {V.}~\bibnamefont {Vadimov}}, \bibinfo {author} {\bibfnamefont {G.}~\bibnamefont {Catelani}}, \bibinfo {author} {\bibfnamefont {V.}~\bibnamefont {Sevriuk}}, \bibinfo {author} {\bibfnamefont {J.}~\bibnamefont {Heinsoo}}, \bibinfo {author} {\bibfnamefont {J.}~\bibnamefont {Räbinä}}, \bibinfo {author} {\bibfnamefont {J.}~\bibnamefont {Ankerhold}},\ and\ \bibinfo {author} {\bibfnamefont
  {M.}~\bibnamefont {Möttönen}},\ }\href {https://arxiv.org/abs/2402.09594} {\bibinfo {title} {Rapid on-demand generation of thermal states in superconducting quantum circuits}} (\bibinfo {year} {2024}),\ \Eprint {https://arxiv.org/abs/2402.09594} {arXiv:2402.09594 [quant-ph]} \BibitemShut {NoStop}%
\bibitem [{\citenamefont {Sandt}\ and\ \citenamefont {Spatschek}(2023)}]{sandt2023efficient}%
  \BibitemOpen
  \bibfield  {author} {\bibinfo {author} {\bibfnamefont {R.}~\bibnamefont {Sandt}}\ and\ \bibinfo {author} {\bibfnamefont {R.}~\bibnamefont {Spatschek}},\ }\bibfield  {title} {\emph {\bibinfo {title} {{Efficient low temperature Monte Carlo sampling using quantum annealing}}},\ }\href@noop {} {\bibfield  {journal} {\bibinfo  {journal} {Scientific Reports}\ }\textbf {\bibinfo {volume} {13}},\ \bibinfo {pages} {6754} (\bibinfo {year} {2023})}\BibitemShut {NoStop}%
\bibitem [{\citenamefont {Wild}\ \emph {et~al.}(2021)\citenamefont {Wild}, \citenamefont {Sels}, \citenamefont {Pichler}, \citenamefont {Zanoci},\ and\ \citenamefont {Lukin}}]{Wild_2021}%
  \BibitemOpen
  \bibfield  {author} {\bibinfo {author} {\bibfnamefont {D.~S.}\ \bibnamefont {Wild}}, \bibinfo {author} {\bibfnamefont {D.}~\bibnamefont {Sels}}, \bibinfo {author} {\bibfnamefont {H.}~\bibnamefont {Pichler}}, \bibinfo {author} {\bibfnamefont {C.}~\bibnamefont {Zanoci}},\ and\ \bibinfo {author} {\bibfnamefont {M.~D.}\ \bibnamefont {Lukin}},\ }\bibfield  {title} {\emph {\bibinfo {title} {Quantum sampling algorithms for near-term devices}},\ }\bibfield  {journal} {\bibinfo  {journal} {Physical Review Letters}\ }\textbf {\bibinfo {volume} {127}},\ \href {https://doi.org/10.1103/physrevlett.127.100504} {10.1103/physrevlett.127.100504} (\bibinfo {year} {2021})\BibitemShut {NoStop}%
\bibitem [{\citenamefont {Holmes}\ \emph {et~al.}(2022)\citenamefont {Holmes}, \citenamefont {Muraleedharan}, \citenamefont {Somma}, \citenamefont {Subasi},\ and\ \citenamefont {{\c{S}}ahino{\u{g}}lu}}]{Holmes2022quantumalgorithms}%
  \BibitemOpen
  \bibfield  {author} {\bibinfo {author} {\bibfnamefont {Z.}~\bibnamefont {Holmes}}, \bibinfo {author} {\bibfnamefont {G.}~\bibnamefont {Muraleedharan}}, \bibinfo {author} {\bibfnamefont {R.~D.}\ \bibnamefont {Somma}}, \bibinfo {author} {\bibfnamefont {Y.}~\bibnamefont {Subasi}},\ and\ \bibinfo {author} {\bibfnamefont {B.}~\bibnamefont {{\c{S}}ahino{\u{g}}lu}},\ }\bibfield  {title} {\emph {\bibinfo {title} {Quantum algorithms from fluctuation theorems: {T}hermal-state preparation}},\ }\href {https://doi.org/10.22331/q-2022-10-06-825} {\bibfield  {journal} {\bibinfo  {journal} {{Quantum}}\ }\textbf {\bibinfo {volume} {6}},\ \bibinfo {pages} {825} (\bibinfo {year} {2022})}\BibitemShut {NoStop}%
\bibitem [{\citenamefont {Poulin}\ and\ \citenamefont {Wocjan}(2009)}]{Poulin_2009}%
  \BibitemOpen
  \bibfield  {author} {\bibinfo {author} {\bibfnamefont {D.}~\bibnamefont {Poulin}}\ and\ \bibinfo {author} {\bibfnamefont {P.}~\bibnamefont {Wocjan}},\ }\bibfield  {title} {\emph {\bibinfo {title} {{Sampling from the Thermal Quantum Gibbs State and Evaluating Partition Functions with a Quantum Computer}}},\ }\bibfield  {journal} {\bibinfo  {journal} {Physical Review Letters}\ }\textbf {\bibinfo {volume} {103}},\ \href {https://doi.org/10.1103/physrevlett.103.220502} {10.1103/physrevlett.103.220502} (\bibinfo {year} {2009})\BibitemShut {NoStop}%
\bibitem [{\citenamefont {Díez-Valle}\ \emph {et~al.}(2023)\citenamefont {Díez-Valle}, \citenamefont {Porras},\ and\ \citenamefont {García-Ripoll}}]{D_ez_Valle_2023_PRL}%
  \BibitemOpen
  \bibfield  {author} {\bibinfo {author} {\bibfnamefont {P.}~\bibnamefont {Díez-Valle}}, \bibinfo {author} {\bibfnamefont {D.}~\bibnamefont {Porras}},\ and\ \bibinfo {author} {\bibfnamefont {J.~J.}\ \bibnamefont {García-Ripoll}},\ }\bibfield  {title} {\emph {\bibinfo {title} {{Quantum Approximate Optimization Algorithm Pseudo-Boltzmann States}}},\ }\bibfield  {journal} {\bibinfo  {journal} {Physical Review Letters}\ }\textbf {\bibinfo {volume} {130}},\ \href {https://doi.org/10.1103/physrevlett.130.050601} {10.1103/physrevlett.130.050601} (\bibinfo {year} {2023})\BibitemShut {NoStop}%
\bibitem [{\citenamefont {Díez-Valle}\ \emph {et~al.}(2024)\citenamefont {Díez-Valle}, \citenamefont {Porras},\ and\ \citenamefont {García-Ripoll}}]{D_ez_Valle_2024}%
  \BibitemOpen
  \bibfield  {author} {\bibinfo {author} {\bibfnamefont {P.}~\bibnamefont {Díez-Valle}}, \bibinfo {author} {\bibfnamefont {D.}~\bibnamefont {Porras}},\ and\ \bibinfo {author} {\bibfnamefont {J.~J.}\ \bibnamefont {García-Ripoll}},\ }\bibfield  {title} {\emph {\bibinfo {title} {Connection between single-layer quantum approximate optimization algorithm interferometry and thermal distribution sampling}},\ }\bibfield  {journal} {\bibinfo  {journal} {Frontiers in Quantum Science and Technology}\ }\textbf {\bibinfo {volume} {3}},\ \href {https://doi.org/10.3389/frqst.2024.1321264} {10.3389/frqst.2024.1321264} (\bibinfo {year} {2024})\BibitemShut {NoStop}%
\bibitem [{\citenamefont {Díez-Valle}\ \emph {et~al.}(2025)\citenamefont {Díez-Valle}, \citenamefont {Gómez-Ruiz}, \citenamefont {Porras},\ and\ \citenamefont {García-Ripoll}}]{diezvalle2025universalresourcesqaoaquantum}%
  \BibitemOpen
  \bibfield  {author} {\bibinfo {author} {\bibfnamefont {P.}~\bibnamefont {Díez-Valle}}, \bibinfo {author} {\bibfnamefont {F.~J.}\ \bibnamefont {Gómez-Ruiz}}, \bibinfo {author} {\bibfnamefont {D.}~\bibnamefont {Porras}},\ and\ \bibinfo {author} {\bibfnamefont {J.~J.}\ \bibnamefont {García-Ripoll}},\ }\href {https://arxiv.org/abs/2506.03241} {\bibinfo {title} {Universal resources for qaoa and quantum annealing}} (\bibinfo {year} {2025}),\ \Eprint {https://arxiv.org/abs/2506.03241} {arXiv:2506.03241 [quant-ph]} \BibitemShut {NoStop}%
\bibitem [{\citenamefont {Lotshaw}\ \emph {et~al.}(2023)\citenamefont {Lotshaw}, \citenamefont {Siopsis}, \citenamefont {Ostrowski}, \citenamefont {Herrman}, \citenamefont {Alam}, \citenamefont {Powers},\ and\ \citenamefont {Humble}}]{PhysRevA.108.042411}%
  \BibitemOpen
  \bibfield  {author} {\bibinfo {author} {\bibfnamefont {P.~C.}\ \bibnamefont {Lotshaw}}, \bibinfo {author} {\bibfnamefont {G.}~\bibnamefont {Siopsis}}, \bibinfo {author} {\bibfnamefont {J.}~\bibnamefont {Ostrowski}}, \bibinfo {author} {\bibfnamefont {R.}~\bibnamefont {Herrman}}, \bibinfo {author} {\bibfnamefont {R.}~\bibnamefont {Alam}}, \bibinfo {author} {\bibfnamefont {S.}~\bibnamefont {Powers}},\ and\ \bibinfo {author} {\bibfnamefont {T.~S.}\ \bibnamefont {Humble}},\ }\bibfield  {title} {\emph {\bibinfo {title} {Approximate boltzmann distributions in quantum approximate optimization}},\ }\href {https://doi.org/10.1103/PhysRevA.108.042411} {\bibfield  {journal} {\bibinfo  {journal} {Phys. Rev. A}\ }\textbf {\bibinfo {volume} {108}},\ \bibinfo {pages} {042411} (\bibinfo {year} {2023})}\BibitemShut {NoStop}%
\bibitem [{\citenamefont {Leontica}\ and\ \citenamefont {Amaro}(2024)}]{Leontica_2024}%
  \BibitemOpen
  \bibfield  {author} {\bibinfo {author} {\bibfnamefont {S.}~\bibnamefont {Leontica}}\ and\ \bibinfo {author} {\bibfnamefont {D.}~\bibnamefont {Amaro}},\ }\bibfield  {title} {\emph {\bibinfo {title} {{Exploring the neighborhood of 1-layer QAOA with instantaneous quantum polynomial circuits}}},\ }\bibfield  {journal} {\bibinfo  {journal} {Physical Review Research}\ }\textbf {\bibinfo {volume} {6}},\ \href {https://doi.org/10.1103/physrevresearch.6.013071} {10.1103/physrevresearch.6.013071} (\bibinfo {year} {2024})\BibitemShut {NoStop}%
\bibitem [{\citenamefont {Abe}\ and\ \citenamefont {Tanaka}(2026)}]{abe2026fairsamplingtemperaturetargetedqaoa}%
  \BibitemOpen
  \bibfield  {author} {\bibinfo {author} {\bibfnamefont {T.}~\bibnamefont {Abe}}\ and\ \bibinfo {author} {\bibfnamefont {S.}~\bibnamefont {Tanaka}},\ }\href {https://arxiv.org/abs/2601.16144} {\bibinfo {title} {{Fair sampling with temperature-targeted QAOA based on quantum-classical correspondence theory}}} (\bibinfo {year} {2026}),\ \Eprint {https://arxiv.org/abs/2601.16144} {arXiv:2601.16144 [quant-ph]} \BibitemShut {NoStop}%
\bibitem [{\citenamefont {Somma}\ \emph {et~al.}(2007)\citenamefont {Somma}, \citenamefont {Batista},\ and\ \citenamefont {Ortiz}}]{PhysRevLett.99.030603}%
  \BibitemOpen
  \bibfield  {author} {\bibinfo {author} {\bibfnamefont {R.~D.}\ \bibnamefont {Somma}}, \bibinfo {author} {\bibfnamefont {C.~D.}\ \bibnamefont {Batista}},\ and\ \bibinfo {author} {\bibfnamefont {G.}~\bibnamefont {Ortiz}},\ }\bibfield  {title} {\emph {\bibinfo {title} {Quantum approach to classical statistical mechanics}},\ }\href {https://doi.org/10.1103/PhysRevLett.99.030603} {\bibfield  {journal} {\bibinfo  {journal} {Phys. Rev. Lett.}\ }\textbf {\bibinfo {volume} {99}},\ \bibinfo {pages} {030603} (\bibinfo {year} {2007})}\BibitemShut {NoStop}%
\bibitem [{\citenamefont {Wu}\ and\ \citenamefont {Hsieh}(2019)}]{Wu_2019}%
  \BibitemOpen
  \bibfield  {author} {\bibinfo {author} {\bibfnamefont {J.}~\bibnamefont {Wu}}\ and\ \bibinfo {author} {\bibfnamefont {T.~H.}\ \bibnamefont {Hsieh}},\ }\bibfield  {title} {\emph {\bibinfo {title} {Variational thermal quantum simulation via thermofield double states}},\ }\bibfield  {journal} {\bibinfo  {journal} {Physical Review Letters}\ }\textbf {\bibinfo {volume} {123}},\ \href {https://doi.org/10.1103/physrevlett.123.220502} {10.1103/physrevlett.123.220502} (\bibinfo {year} {2019})\BibitemShut {NoStop}%
\bibitem [{\citenamefont {Warren}\ \emph {et~al.}(2022)\citenamefont {Warren}, \citenamefont {Zhu}, \citenamefont {Mayhall}, \citenamefont {Barnes},\ and\ \citenamefont {Economou}}]{warren2022adaptivevariationalalgorithmsquantum}%
  \BibitemOpen
  \bibfield  {author} {\bibinfo {author} {\bibfnamefont {A.}~\bibnamefont {Warren}}, \bibinfo {author} {\bibfnamefont {L.}~\bibnamefont {Zhu}}, \bibinfo {author} {\bibfnamefont {N.~J.}\ \bibnamefont {Mayhall}}, \bibinfo {author} {\bibfnamefont {E.}~\bibnamefont {Barnes}},\ and\ \bibinfo {author} {\bibfnamefont {S.~E.}\ \bibnamefont {Economou}},\ }\href {https://arxiv.org/abs/2203.12757} {\bibinfo {title} {Adaptive variational algorithms for quantum gibbs state preparation}} (\bibinfo {year} {2022}),\ \Eprint {https://arxiv.org/abs/2203.12757} {arXiv:2203.12757 [quant-ph]} \BibitemShut {NoStop}%
\bibitem [{\citenamefont {Nakano}\ \emph {et~al.}(2025)\citenamefont {Nakano}, \citenamefont {Okada},\ and\ \citenamefont {Fujii}}]{nakano2025neuralnetworkassistedmontecarlosampling}%
  \BibitemOpen
  \bibfield  {author} {\bibinfo {author} {\bibfnamefont {Y.}~\bibnamefont {Nakano}}, \bibinfo {author} {\bibfnamefont {K.~N.}\ \bibnamefont {Okada}},\ and\ \bibinfo {author} {\bibfnamefont {K.}~\bibnamefont {Fujii}},\ }\href {https://arxiv.org/abs/2506.01335} {\bibinfo {title} {{Neural-network-assisted Monte Carlo sampling trained by Quantum Approximate Optimization Algorithm}}} (\bibinfo {year} {2025}),\ \Eprint {https://arxiv.org/abs/2506.01335} {arXiv:2506.01335 [quant-ph]} \BibitemShut {NoStop}%
\bibitem [{\citenamefont {Metropolis}\ \emph {et~al.}(1953)\citenamefont {Metropolis}, \citenamefont {Rosenbluth}, \citenamefont {Rosenbluth}, \citenamefont {Teller},\ and\ \citenamefont {Teller}}]{metropolis1953equation}%
  \BibitemOpen
  \bibfield  {author} {\bibinfo {author} {\bibfnamefont {N.}~\bibnamefont {Metropolis}}, \bibinfo {author} {\bibfnamefont {A.~W.}\ \bibnamefont {Rosenbluth}}, \bibinfo {author} {\bibfnamefont {M.~N.}\ \bibnamefont {Rosenbluth}}, \bibinfo {author} {\bibfnamefont {A.~H.}\ \bibnamefont {Teller}},\ and\ \bibinfo {author} {\bibfnamefont {E.}~\bibnamefont {Teller}},\ }\bibfield  {title} {\emph {\bibinfo {title} {Equation of state calculations by fast computing machines}},\ }\href@noop {} {\bibfield  {journal} {\bibinfo  {journal} {The journal of chemical physics}\ }\textbf {\bibinfo {volume} {21}},\ \bibinfo {pages} {1087--1092} (\bibinfo {year} {1953})}\BibitemShut {NoStop}%
\bibitem [{\citenamefont {Layden}\ \emph {et~al.}(2023)\citenamefont {Layden}, \citenamefont {Mazzola}, \citenamefont {Mishmash}, \citenamefont {Motta}, \citenamefont {Wocjan}, \citenamefont {Kim},\ and\ \citenamefont {Sheldon}}]{Layden_2023}%
  \BibitemOpen
  \bibfield  {author} {\bibinfo {author} {\bibfnamefont {D.}~\bibnamefont {Layden}}, \bibinfo {author} {\bibfnamefont {G.}~\bibnamefont {Mazzola}}, \bibinfo {author} {\bibfnamefont {R.~V.}\ \bibnamefont {Mishmash}}, \bibinfo {author} {\bibfnamefont {M.}~\bibnamefont {Motta}}, \bibinfo {author} {\bibfnamefont {P.}~\bibnamefont {Wocjan}}, \bibinfo {author} {\bibfnamefont {J.-S.}\ \bibnamefont {Kim}},\ and\ \bibinfo {author} {\bibfnamefont {S.}~\bibnamefont {Sheldon}},\ }\bibfield  {title} {\emph {\bibinfo {title} {{Quantum-enhanced Markov chain Monte Carlo}}},\ }\href {https://doi.org/10.1038/s41586-023-06095-4} {\bibfield  {journal} {\bibinfo  {journal} {Nature}\ }\textbf {\bibinfo {volume} {619}},\ \bibinfo {pages} {282–287} (\bibinfo {year} {2023})}\BibitemShut {NoStop}%
\bibitem [{\citenamefont {Nakano}\ \emph {et~al.}(2024)\citenamefont {Nakano}, \citenamefont {Hakoshima}, \citenamefont {Mitarai},\ and\ \citenamefont {Fujii}}]{Nakano_2024}%
  \BibitemOpen
  \bibfield  {author} {\bibinfo {author} {\bibfnamefont {Y.}~\bibnamefont {Nakano}}, \bibinfo {author} {\bibfnamefont {H.}~\bibnamefont {Hakoshima}}, \bibinfo {author} {\bibfnamefont {K.}~\bibnamefont {Mitarai}},\ and\ \bibinfo {author} {\bibfnamefont {K.}~\bibnamefont {Fujii}},\ }\bibfield  {title} {\emph {\bibinfo {title} {{Markov-chain Monte Carlo method enhanced by a quantum alternating operator ansatz}}},\ }\bibfield  {journal} {\bibinfo  {journal} {Physical Review Research}\ }\textbf {\bibinfo {volume} {6}},\ \href {https://doi.org/10.1103/physrevresearch.6.033105} {10.1103/physrevresearch.6.033105} (\bibinfo {year} {2024})\BibitemShut {NoStop}%
\bibitem [{\citenamefont {Sherrington}\ and\ \citenamefont {Kirkpatrick}(1975)}]{PhysRevLett.35.1792}%
  \BibitemOpen
  \bibfield  {author} {\bibinfo {author} {\bibfnamefont {D.}~\bibnamefont {Sherrington}}\ and\ \bibinfo {author} {\bibfnamefont {S.}~\bibnamefont {Kirkpatrick}},\ }\bibfield  {title} {\emph {\bibinfo {title} {{Solvable Model of a Spin-Glass}}},\ }\href {https://doi.org/10.1103/PhysRevLett.35.1792} {\bibfield  {journal} {\bibinfo  {journal} {Phys. Rev. Lett.}\ }\textbf {\bibinfo {volume} {35}},\ \bibinfo {pages} {1792--1796} (\bibinfo {year} {1975})}\BibitemShut {NoStop}%
\bibitem [{\citenamefont {Akshay}\ \emph {et~al.}(2021)\citenamefont {Akshay}, \citenamefont {Rabinovich}, \citenamefont {Campos},\ and\ \citenamefont {Biamonte}}]{PhysRevA.104.L010401}%
  \BibitemOpen
  \bibfield  {author} {\bibinfo {author} {\bibfnamefont {V.}~\bibnamefont {Akshay}}, \bibinfo {author} {\bibfnamefont {D.}~\bibnamefont {Rabinovich}}, \bibinfo {author} {\bibfnamefont {E.}~\bibnamefont {Campos}},\ and\ \bibinfo {author} {\bibfnamefont {J.}~\bibnamefont {Biamonte}},\ }\bibfield  {title} {\emph {\bibinfo {title} {{Parameter Concentration in Quantum Approximate Optimization}}},\ }\href {https://doi.org/10.1103/PhysRevA.104.L010401} {\bibfield  {journal} {\bibinfo  {journal} {Physical Review A}\ }\textbf {\bibinfo {volume} {104}},\ \bibinfo {pages} {L010401} (\bibinfo {year} {2021})},\ \Eprint {https://arxiv.org/abs/2103.11976} {arXiv:2103.11976} \BibitemShut {NoStop}%
\bibitem [{\citenamefont {Brandao}\ \emph {et~al.}(2018)\citenamefont {Brandao}, \citenamefont {Broughton}, \citenamefont {Farhi}, \citenamefont {Gutmann},\ and\ \citenamefont {Neven}}]{brandao2018fixedcontrolparametersquantum}%
  \BibitemOpen
  \bibfield  {author} {\bibinfo {author} {\bibfnamefont {F.~G. S.~L.}\ \bibnamefont {Brandao}}, \bibinfo {author} {\bibfnamefont {M.}~\bibnamefont {Broughton}}, \bibinfo {author} {\bibfnamefont {E.}~\bibnamefont {Farhi}}, \bibinfo {author} {\bibfnamefont {S.}~\bibnamefont {Gutmann}},\ and\ \bibinfo {author} {\bibfnamefont {H.}~\bibnamefont {Neven}},\ }\bibfield  {title} {\emph {\bibinfo {title} {{For Fixed Control Parameters the Quantum Approximate Optimization Algorithm's Objective Function Value Concentrates for Typical Instances}}},\ }\href@noop {} {\bibfield  {journal} {\bibinfo  {journal} {arXiv preprint}\ } (\bibinfo {year} {2018})},\ \Eprint {https://arxiv.org/abs/1812.04170} {arXiv:1812.04170} \BibitemShut {NoStop}%
\bibitem [{\citenamefont {Boulebnane}\ and\ \citenamefont {Montanaro}(2021)}]{boulebnane2021predictingparametersquantumapproximate}%
  \BibitemOpen
  \bibfield  {author} {\bibinfo {author} {\bibfnamefont {S.}~\bibnamefont {Boulebnane}}\ and\ \bibinfo {author} {\bibfnamefont {A.}~\bibnamefont {Montanaro}},\ }\bibfield  {title} {\emph {\bibinfo {title} {Predicting parameters for the quantum approximate optimization algorithm for max-cut from the infinite-size limit}},\ }\href@noop {} {\bibfield  {journal} {\bibinfo  {journal} {arXiv preprint}\ } (\bibinfo {year} {2021})},\ \Eprint {https://arxiv.org/abs/2110.10685} {arXiv:2110.10685} \BibitemShut {NoStop}%
\bibitem [{\citenamefont {Shaydulin}\ \emph {et~al.}(2023)\citenamefont {Shaydulin}, \citenamefont {Lotshaw}, \citenamefont {Larson}, \citenamefont {Ostrowski},\ and\ \citenamefont {Humble}}]{Shaydulin_2023}%
  \BibitemOpen
  \bibfield  {author} {\bibinfo {author} {\bibfnamefont {R.}~\bibnamefont {Shaydulin}}, \bibinfo {author} {\bibfnamefont {P.~C.}\ \bibnamefont {Lotshaw}}, \bibinfo {author} {\bibfnamefont {J.}~\bibnamefont {Larson}}, \bibinfo {author} {\bibfnamefont {J.}~\bibnamefont {Ostrowski}},\ and\ \bibinfo {author} {\bibfnamefont {T.~S.}\ \bibnamefont {Humble}},\ }\bibfield  {title} {\emph {\bibinfo {title} {{Parameter Transfer for Quantum Approximate Optimization of Weighted MaxCut}}},\ }\href {https://doi.org/10.1145/3584706} {\bibfield  {journal} {\bibinfo  {journal} {ACM Transactions on Quantum Computing}\ }\textbf {\bibinfo {volume} {4}},\ \bibinfo {pages} {19} (\bibinfo {year} {2023})},\ \Eprint {https://arxiv.org/abs/2201.11785} {arXiv:2201.11785} \BibitemShut {NoStop}%
\bibitem [{\citenamefont {Katial}\ \emph {et~al.}(2024)\citenamefont {Katial}, \citenamefont {Smith-Miles},\ and\ \citenamefont {Hill}}]{katial2024instancedependenceoptimalparameters}%
  \BibitemOpen
  \bibfield  {author} {\bibinfo {author} {\bibfnamefont {V.}~\bibnamefont {Katial}}, \bibinfo {author} {\bibfnamefont {K.}~\bibnamefont {Smith-Miles}},\ and\ \bibinfo {author} {\bibfnamefont {C.}~\bibnamefont {Hill}},\ }\href {https://arxiv.org/abs/2401.08142} {\bibinfo {title} {{On the Instance Dependence of Optimal Parameters for the Quantum Approximate Optimisation Algorithm: Insights via Instance Space Analysis}}} (\bibinfo {year} {2024}),\ \Eprint {https://arxiv.org/abs/2401.08142} {arXiv:2401.08142 [quant-ph]} \BibitemShut {NoStop}%
\bibitem [{\citenamefont {Basso}\ \emph {et~al.}(2022)\citenamefont {Basso}, \citenamefont {Farhi}, \citenamefont {Marwaha}, \citenamefont {Villalonga},\ and\ \citenamefont {Zhou}}]{LIPICS.TQC.2022.7}%
  \BibitemOpen
  \bibfield  {author} {\bibinfo {author} {\bibfnamefont {J.}~\bibnamefont {Basso}}, \bibinfo {author} {\bibfnamefont {E.}~\bibnamefont {Farhi}}, \bibinfo {author} {\bibfnamefont {K.}~\bibnamefont {Marwaha}}, \bibinfo {author} {\bibfnamefont {B.}~\bibnamefont {Villalonga}},\ and\ \bibinfo {author} {\bibfnamefont {L.}~\bibnamefont {Zhou}},\ }in\ \href {https://doi.org/10.4230/LIPICS.TQC.2022.7} {\emph {\bibinfo {booktitle} {17th Conference on the Theory of Quantum Computation, Communication and Cryptography TQC'22}}}\ (\bibinfo {year} {2022})\ \Eprint {https://arxiv.org/abs/2110.14206} {arXiv:2110.14206} \BibitemShut {NoStop}%
\bibitem [{\citenamefont {Sakai}\ \emph {et~al.}(2024)\citenamefont {Sakai}, \citenamefont {Matsuyama}, \citenamefont {Tam}, \citenamefont {Yamashiro},\ and\ \citenamefont {Fujii}}]{sakai2024linearlysimplifiedqaoaparameters}%
  \BibitemOpen
  \bibfield  {author} {\bibinfo {author} {\bibfnamefont {R.}~\bibnamefont {Sakai}}, \bibinfo {author} {\bibfnamefont {H.}~\bibnamefont {Matsuyama}}, \bibinfo {author} {\bibfnamefont {W.-H.}\ \bibnamefont {Tam}}, \bibinfo {author} {\bibfnamefont {Y.}~\bibnamefont {Yamashiro}},\ and\ \bibinfo {author} {\bibfnamefont {K.}~\bibnamefont {Fujii}},\ }\bibfield  {title} {\emph {\bibinfo {title} {{Linearly simplified QAOA parameters and transferability}}},\ }\href@noop {} {\bibfield  {journal} {\bibinfo  {journal} {arXiv preprint}\ } (\bibinfo {year} {2024})},\ \Eprint {https://arxiv.org/abs/2405.00655} {arXiv:2405.00655} \BibitemShut {NoStop}%
\bibitem [{\citenamefont {Farhi}\ \emph {et~al.}(2022)\citenamefont {Farhi}, \citenamefont {Goldstone}, \citenamefont {Gutmann},\ and\ \citenamefont {Zhou}}]{Farhi_2022}%
  \BibitemOpen
  \bibfield  {author} {\bibinfo {author} {\bibfnamefont {E.}~\bibnamefont {Farhi}}, \bibinfo {author} {\bibfnamefont {J.}~\bibnamefont {Goldstone}}, \bibinfo {author} {\bibfnamefont {S.}~\bibnamefont {Gutmann}},\ and\ \bibinfo {author} {\bibfnamefont {L.}~\bibnamefont {Zhou}},\ }\bibfield  {title} {\emph {\bibinfo {title} {{The Quantum Approximate Optimization Algorithm and the Sherrington-Kirkpatrick Model at Infinite Size}}},\ }\href {https://doi.org/10.22331/q-2022-07-07-759} {\bibfield  {journal} {\bibinfo  {journal} {Quantum}\ }\textbf {\bibinfo {volume} {6}},\ \bibinfo {pages} {759} (\bibinfo {year} {2022})},\ \Eprint {https://arxiv.org/abs/1910.08187} {arXiv:1910.08187} \BibitemShut {NoStop}%
\bibitem [{\citenamefont {Galda}\ \emph {et~al.}(2021)\citenamefont {Galda}, \citenamefont {Liu}, \citenamefont {Lykov}, \citenamefont {Alexeev},\ and\ \citenamefont {Safro}}]{galda2021transferabilityoptimalqaoaparameters}%
  \BibitemOpen
  \bibfield  {author} {\bibinfo {author} {\bibfnamefont {A.}~\bibnamefont {Galda}}, \bibinfo {author} {\bibfnamefont {X.}~\bibnamefont {Liu}}, \bibinfo {author} {\bibfnamefont {D.}~\bibnamefont {Lykov}}, \bibinfo {author} {\bibfnamefont {Y.}~\bibnamefont {Alexeev}},\ and\ \bibinfo {author} {\bibfnamefont {I.}~\bibnamefont {Safro}},\ }\href {https://arxiv.org/abs/2106.07531} {\bibinfo {title} {{Transferability of optimal QAOA parameters between random graphs}}} (\bibinfo {year} {2021}),\ \Eprint {https://arxiv.org/abs/2106.07531} {arXiv:2106.07531 [quant-ph]} \BibitemShut {NoStop}%
\bibitem [{\citenamefont {Galda}\ \emph {et~al.}(2023)\citenamefont {Galda}, \citenamefont {Gupta}, \citenamefont {Falla}, \citenamefont {Liu}, \citenamefont {Lykov}, \citenamefont {Alexeev},\ and\ \citenamefont {Safro}}]{galda2023similaritybasedparametertransferabilityquantum}%
  \BibitemOpen
  \bibfield  {author} {\bibinfo {author} {\bibfnamefont {A.}~\bibnamefont {Galda}}, \bibinfo {author} {\bibfnamefont {E.}~\bibnamefont {Gupta}}, \bibinfo {author} {\bibfnamefont {J.}~\bibnamefont {Falla}}, \bibinfo {author} {\bibfnamefont {X.}~\bibnamefont {Liu}}, \bibinfo {author} {\bibfnamefont {D.}~\bibnamefont {Lykov}}, \bibinfo {author} {\bibfnamefont {Y.}~\bibnamefont {Alexeev}},\ and\ \bibinfo {author} {\bibfnamefont {I.}~\bibnamefont {Safro}},\ }\href {https://arxiv.org/abs/2307.05420} {\bibinfo {title} {{Similarity-Based Parameter Transferability in the Quantum Approximate Optimization Algorithm}}} (\bibinfo {year} {2023}),\ \Eprint {https://arxiv.org/abs/2307.05420} {arXiv:2307.05420 [quant-ph]} \BibitemShut {NoStop}%
\bibitem [{\citenamefont {Chernyavskiy}\ \emph {et~al.}(2025)\citenamefont {Chernyavskiy}, \citenamefont {Kulikov}, \citenamefont {Bantysh}, \citenamefont {Bogdanov}, \citenamefont {Fedorov},\ and\ \citenamefont {Kiktenko}}]{chernyavskiy2025improvingqaoaapproximatequbo}%
  \BibitemOpen
  \bibfield  {author} {\bibinfo {author} {\bibfnamefont {A.~Y.}\ \bibnamefont {Chernyavskiy}}, \bibinfo {author} {\bibfnamefont {D.~A.}\ \bibnamefont {Kulikov}}, \bibinfo {author} {\bibfnamefont {B.~I.}\ \bibnamefont {Bantysh}}, \bibinfo {author} {\bibfnamefont {Y.~I.}\ \bibnamefont {Bogdanov}}, \bibinfo {author} {\bibfnamefont {A.~K.}\ \bibnamefont {Fedorov}},\ and\ \bibinfo {author} {\bibfnamefont {E.~O.}\ \bibnamefont {Kiktenko}},\ }\href {https://arxiv.org/abs/2509.19035} {\bibinfo {title} {{Improving QAOA to find approximate QUBO solutions in O(1) shots}}} (\bibinfo {year} {2025}),\ \Eprint {https://arxiv.org/abs/2509.19035} {arXiv:2509.19035 [quant-ph]} \BibitemShut {NoStop}%
\bibitem [{\citenamefont {Wurtz}\ and\ \citenamefont {Lykov}(2021)}]{PhysRevA.104.052419}%
  \BibitemOpen
  \bibfield  {author} {\bibinfo {author} {\bibfnamefont {J.}~\bibnamefont {Wurtz}}\ and\ \bibinfo {author} {\bibfnamefont {D.}~\bibnamefont {Lykov}},\ }\bibfield  {title} {\emph {\bibinfo {title} {{Fixed-angle conjectures for the quantum approximate optimization algorithm on regular MaxCut graphs}}},\ }\href {https://doi.org/10.1103/PhysRevA.104.052419} {\bibfield  {journal} {\bibinfo  {journal} {Phys. Rev. A}\ }\textbf {\bibinfo {volume} {104}},\ \bibinfo {pages} {052419} (\bibinfo {year} {2021})}\BibitemShut {NoStop}%
\bibitem [{\citenamefont {Rudolph}\ \emph {et~al.}(2021)\citenamefont {Rudolph}, \citenamefont {Sim}, \citenamefont {Raza}, \citenamefont {Stechly}, \citenamefont {McClean}, \citenamefont {Anschuetz}, \citenamefont {Serrano},\ and\ \citenamefont {Perdomo-Ortiz}}]{rudolph2021orqvizvisualizinghighdimensionallandscapes}%
  \BibitemOpen
  \bibfield  {author} {\bibinfo {author} {\bibfnamefont {M.~S.}\ \bibnamefont {Rudolph}}, \bibinfo {author} {\bibfnamefont {S.}~\bibnamefont {Sim}}, \bibinfo {author} {\bibfnamefont {A.}~\bibnamefont {Raza}}, \bibinfo {author} {\bibfnamefont {M.}~\bibnamefont {Stechly}}, \bibinfo {author} {\bibfnamefont {J.~R.}\ \bibnamefont {McClean}}, \bibinfo {author} {\bibfnamefont {E.~R.}\ \bibnamefont {Anschuetz}}, \bibinfo {author} {\bibfnamefont {L.}~\bibnamefont {Serrano}},\ and\ \bibinfo {author} {\bibfnamefont {A.}~\bibnamefont {Perdomo-Ortiz}},\ }\href {https://arxiv.org/abs/2111.04695} {\bibinfo {title} {{ORQVIZ: Visualizing High-Dimensional Landscapes in Variational Quantum Algorithms}}} (\bibinfo {year} {2021}),\ \Eprint {https://arxiv.org/abs/2111.04695} {arXiv:2111.04695 [quant-ph]} \BibitemShut {NoStop}%
\bibitem [{\citenamefont {Krüger}\ and\ \citenamefont {Mauerer}(2025)}]{Kr_ger_2025}%
  \BibitemOpen
  \bibfield  {author} {\bibinfo {author} {\bibfnamefont {T.}~\bibnamefont {Krüger}}\ and\ \bibinfo {author} {\bibfnamefont {W.}~\bibnamefont {Mauerer}},\ }\bibfield  {title} {\emph {\bibinfo {title} {{Out of the Loop: Structural Approximation of Optimisation Landscapes and non-Iterative Quantum Optimisation}}},\ }\href {https://doi.org/10.22331/q-2025-11-06-1903} {\bibfield  {journal} {\bibinfo  {journal} {Quantum}\ }\textbf {\bibinfo {volume} {9}},\ \bibinfo {pages} {1903} (\bibinfo {year} {2025})}\BibitemShut {NoStop}%
\bibitem [{\citenamefont {Claes}\ and\ \citenamefont {Dam}(2021)}]{Claes_2021}%
  \BibitemOpen
  \bibfield  {author} {\bibinfo {author} {\bibfnamefont {J.}~\bibnamefont {Claes}}\ and\ \bibinfo {author} {\bibfnamefont {W.~v.}\ \bibnamefont {Dam}},\ }\bibfield  {title} {\emph {\bibinfo {title} {{Instance Independence of Single Layer Quantum Approximate Optimization Algorithm on Mixed-Spin Models at Infinite Size}}},\ }\href {https://doi.org/10.22331/q-2021-09-15-542} {\bibfield  {journal} {\bibinfo  {journal} {Quantum}\ }\textbf {\bibinfo {volume} {5}},\ \bibinfo {pages} {542} (\bibinfo {year} {2021})}\BibitemShut {NoStop}%
\bibitem [{\citenamefont {Lykov}\ \emph {et~al.}(2023)\citenamefont {Lykov}, \citenamefont {Wurtz}, \citenamefont {Poole}, \citenamefont {Saffman}, \citenamefont {Noel},\ and\ \citenamefont {Alexeev}}]{Lykov_2023}%
  \BibitemOpen
  \bibfield  {author} {\bibinfo {author} {\bibfnamefont {D.}~\bibnamefont {Lykov}}, \bibinfo {author} {\bibfnamefont {J.}~\bibnamefont {Wurtz}}, \bibinfo {author} {\bibfnamefont {C.}~\bibnamefont {Poole}}, \bibinfo {author} {\bibfnamefont {M.}~\bibnamefont {Saffman}}, \bibinfo {author} {\bibfnamefont {T.}~\bibnamefont {Noel}},\ and\ \bibinfo {author} {\bibfnamefont {Y.}~\bibnamefont {Alexeev}},\ }\bibfield  {title} {\emph {\bibinfo {title} {Sampling frequency thresholds for the quantum advantage of the quantum approximate optimization algorithm}},\ }\bibfield  {journal} {\bibinfo  {journal} {npj Quantum Information}\ }\textbf {\bibinfo {volume} {9}},\ \href {https://doi.org/10.1038/s41534-023-00718-4} {10.1038/s41534-023-00718-4} (\bibinfo {year} {2023})\BibitemShut {NoStop}%
\bibitem [{\citenamefont {Bezanson}\ \emph {et~al.}(2017)\citenamefont {Bezanson}, \citenamefont {Edelman}, \citenamefont {Karpinski},\ and\ \citenamefont {Shah}}]{bezanson2017julia}%
  \BibitemOpen
  \bibfield  {author} {\bibinfo {author} {\bibfnamefont {J.}~\bibnamefont {Bezanson}}, \bibinfo {author} {\bibfnamefont {A.}~\bibnamefont {Edelman}}, \bibinfo {author} {\bibfnamefont {S.}~\bibnamefont {Karpinski}},\ and\ \bibinfo {author} {\bibfnamefont {V.~B.}\ \bibnamefont {Shah}},\ }\bibfield  {title} {\emph {\bibinfo {title} {{Julia: A Fresh Approach to Numerical Computing}}},\ }\href {https://doi.org/10.1137/141000671} {\bibfield  {journal} {\bibinfo  {journal} {SIAM review}\ }\textbf {\bibinfo {volume} {59}},\ \bibinfo {pages} {65–98} (\bibinfo {year} {2017})},\ \Eprint {https://arxiv.org/abs/1411.1607} {arXiv:1411.1607} \BibitemShut {NoStop}%
\bibitem [{\citenamefont {Golden}\ \emph {et~al.}(2023)\citenamefont {Golden}, \citenamefont {Baertschi}, \citenamefont {O'Malley}, \citenamefont {Pelofske},\ and\ \citenamefont {Eidenbenz}}]{Golden_2023}%
  \BibitemOpen
  \bibfield  {author} {\bibinfo {author} {\bibfnamefont {J.}~\bibnamefont {Golden}}, \bibinfo {author} {\bibfnamefont {A.}~\bibnamefont {Baertschi}}, \bibinfo {author} {\bibfnamefont {D.}~\bibnamefont {O'Malley}}, \bibinfo {author} {\bibfnamefont {E.}~\bibnamefont {Pelofske}},\ and\ \bibinfo {author} {\bibfnamefont {S.}~\bibnamefont {Eidenbenz}},\ }in\ \href {https://doi.org/10.1145/3624062.3624220} {\emph {\bibinfo {booktitle} {Workshops of The International Conference on High Performance Computing, Network, Storage, and Analysis SC-W'23}}}\ (\bibinfo  {publisher} {ACM},\ \bibinfo {year} {2023})\ p.\ \bibinfo {pages} {1454–1459},\ \Eprint {https://arxiv.org/abs/2312.06451} {arXiv:2312.06451} \BibitemShut {NoStop}%
\bibitem [{\citenamefont {Virtanen}\ \emph {et~al.}(2020)\citenamefont {Virtanen}, \citenamefont {Gommers}, \citenamefont {Oliphant}, \citenamefont {Haberland}, \citenamefont {Reddy}, \citenamefont {Cournapeau}, \citenamefont {Burovski}, \citenamefont {Peterson}, \citenamefont {Weckesser}, \citenamefont {Bright}, \citenamefont {{van der Walt}}, \citenamefont {Brett}, \citenamefont {Wilson}, \citenamefont {Millman}, \citenamefont {Mayorov}, \citenamefont {Nelson}, \citenamefont {Jones}, \citenamefont {Kern}, \citenamefont {Larson}, \citenamefont {Carey}, \citenamefont {Polat}, \citenamefont {Feng}, \citenamefont {Moore}, \citenamefont {{VanderPlas}}, \citenamefont {Laxalde}, \citenamefont {Perktold}, \citenamefont {Cimrman}, \citenamefont {Henriksen}, \citenamefont {Quintero}, \citenamefont {Harris}, \citenamefont {Archibald}, \citenamefont {Ribeiro}, \citenamefont {Pedregosa}, \citenamefont {{van Mulbregt}},\ and\ \citenamefont {{SciPy 1.0 Contributors}}}]{2020SciPy-NMeth}%
  \BibitemOpen
  \bibfield  {author} {\bibinfo {author} {\bibfnamefont {P.}~\bibnamefont {Virtanen}}, \bibinfo {author} {\bibfnamefont {R.}~\bibnamefont {Gommers}}, \bibinfo {author} {\bibfnamefont {T.~E.}\ \bibnamefont {Oliphant}}, \bibinfo {author} {\bibfnamefont {M.}~\bibnamefont {Haberland}}, \bibinfo {author} {\bibfnamefont {T.}~\bibnamefont {Reddy}}, \bibinfo {author} {\bibfnamefont {D.}~\bibnamefont {Cournapeau}}, \bibinfo {author} {\bibfnamefont {E.}~\bibnamefont {Burovski}}, \bibinfo {author} {\bibfnamefont {P.}~\bibnamefont {Peterson}}, \bibinfo {author} {\bibfnamefont {W.}~\bibnamefont {Weckesser}}, \bibinfo {author} {\bibfnamefont {J.}~\bibnamefont {Bright}}, \bibinfo {author} {\bibfnamefont {S.~J.}\ \bibnamefont {{van der Walt}}}, \bibinfo {author} {\bibfnamefont {M.}~\bibnamefont {Brett}}, \bibinfo {author} {\bibfnamefont {J.}~\bibnamefont {Wilson}}, \bibinfo {author} {\bibfnamefont {K.~J.}\ \bibnamefont {Millman}}, \bibinfo {author} {\bibfnamefont {N.}~\bibnamefont {Mayorov}}, \bibinfo {author} {\bibfnamefont
  {A.~R.~J.}\ \bibnamefont {Nelson}}, \bibinfo {author} {\bibfnamefont {E.}~\bibnamefont {Jones}}, \bibinfo {author} {\bibfnamefont {R.}~\bibnamefont {Kern}}, \bibinfo {author} {\bibfnamefont {E.}~\bibnamefont {Larson}}, \bibinfo {author} {\bibfnamefont {C.~J.}\ \bibnamefont {Carey}}, \bibinfo {author} {\bibfnamefont {{\.I}.}~\bibnamefont {Polat}}, \bibinfo {author} {\bibfnamefont {Y.}~\bibnamefont {Feng}}, \bibinfo {author} {\bibfnamefont {E.~W.}\ \bibnamefont {Moore}}, \bibinfo {author} {\bibfnamefont {J.}~\bibnamefont {{VanderPlas}}}, \bibinfo {author} {\bibfnamefont {D.}~\bibnamefont {Laxalde}}, \bibinfo {author} {\bibfnamefont {J.}~\bibnamefont {Perktold}}, \bibinfo {author} {\bibfnamefont {R.}~\bibnamefont {Cimrman}}, \bibinfo {author} {\bibfnamefont {I.}~\bibnamefont {Henriksen}}, \bibinfo {author} {\bibfnamefont {E.~A.}\ \bibnamefont {Quintero}}, \bibinfo {author} {\bibfnamefont {C.~R.}\ \bibnamefont {Harris}}, \bibinfo {author} {\bibfnamefont {A.~M.}\ \bibnamefont {Archibald}}, \bibinfo {author}
  {\bibfnamefont {A.~H.}\ \bibnamefont {Ribeiro}}, \bibinfo {author} {\bibfnamefont {F.}~\bibnamefont {Pedregosa}}, \bibinfo {author} {\bibfnamefont {P.}~\bibnamefont {{van Mulbregt}}},\ and\ \bibinfo {author} {\bibnamefont {{SciPy 1.0 Contributors}}},\ }\bibfield  {title} {\emph {\bibinfo {title} {{{SciPy} 1.0: Fundamental Algorithms for Scientific Computing in Python}}},\ }\href {https://doi.org/10.1038/s41592-019-0686-2} {\bibfield  {journal} {\bibinfo  {journal} {Nature Methods}\ }\textbf {\bibinfo {volume} {17}},\ \bibinfo {pages} {261--272} (\bibinfo {year} {2020})}\BibitemShut {NoStop}%
\bibitem [{\citenamefont {Gao}\ and\ \citenamefont {Han}(2012)}]{10.1007/s10589-010-9329-3}%
  \BibitemOpen
  \bibfield  {author} {\bibinfo {author} {\bibfnamefont {F.}~\bibnamefont {Gao}}\ and\ \bibinfo {author} {\bibfnamefont {L.}~\bibnamefont {Han}},\ }\bibfield  {title} {\emph {\bibinfo {title} {Implementing the nelder-mead simplex algorithm with adaptive parameters}},\ }\href {https://doi.org/10.1007/s10589-010-9329-3} {\bibfield  {journal} {\bibinfo  {journal} {Comput. Optim. Appl.}\ }\textbf {\bibinfo {volume} {51}},\ \bibinfo {pages} {259–277} (\bibinfo {year} {2012})}\BibitemShut {NoStop}%
\bibitem [{\citenamefont {Nocedal}\ and\ \citenamefont {Wright}(2006)}]{nocedal2006numerical}%
  \BibitemOpen
  \bibfield  {author} {\bibinfo {author} {\bibfnamefont {J.}~\bibnamefont {Nocedal}}\ and\ \bibinfo {author} {\bibfnamefont {S.~J.}\ \bibnamefont {Wright}},\ }\href@noop {} {\emph {\bibinfo {title} {Numerical optimization}}}\ (\bibinfo  {publisher} {Springer},\ \bibinfo {year} {2006})\BibitemShut {NoStop}%
\bibitem [{\citenamefont {Conn}\ \emph {et~al.}(2000)\citenamefont {Conn}, \citenamefont {Gould},\ and\ \citenamefont {Toint}}]{conn2000trust}%
  \BibitemOpen
  \bibfield  {author} {\bibinfo {author} {\bibfnamefont {A.~R.}\ \bibnamefont {Conn}}, \bibinfo {author} {\bibfnamefont {N.~I.}\ \bibnamefont {Gould}},\ and\ \bibinfo {author} {\bibfnamefont {P.~L.}\ \bibnamefont {Toint}},\ }\href@noop {} {\emph {\bibinfo {title} {Trust region methods}}}\ (\bibinfo  {publisher} {SIAM},\ \bibinfo {year} {2000})\BibitemShut {NoStop}%
\bibitem [{\citenamefont {Lam}\ \emph {et~al.}(2015)\citenamefont {Lam}, \citenamefont {Pitrou},\ and\ \citenamefont {Seibert}}]{10.1145/2833157.2833162}%
  \BibitemOpen
  \bibfield  {author} {\bibinfo {author} {\bibfnamefont {S.~K.}\ \bibnamefont {Lam}}, \bibinfo {author} {\bibfnamefont {A.}~\bibnamefont {Pitrou}},\ and\ \bibinfo {author} {\bibfnamefont {S.}~\bibnamefont {Seibert}},\ }in\ \href {https://doi.org/10.1145/2833157.2833162} {\emph {\bibinfo {booktitle} {Proceedings of the Second Workshop on the LLVM Compiler Infrastructure in HPC}}},\ \bibinfo {series and number} {LLVM '15}\ (\bibinfo  {publisher} {Association for Computing Machinery},\ \bibinfo {address} {New York, NY, USA},\ \bibinfo {year} {2015})\BibitemShut {NoStop}%
\bibitem [{\citenamefont {Harris}\ \emph {et~al.}(2020)\citenamefont {Harris}, \citenamefont {Millman}, \citenamefont {van~der Walt}, \citenamefont {Gommers}, \citenamefont {Virtanen}, \citenamefont {Cournapeau}, \citenamefont {Wieser}, \citenamefont {Taylor}, \citenamefont {Berg}, \citenamefont {Smith}, \citenamefont {Kern}, \citenamefont {Picus}, \citenamefont {Hoyer}, \citenamefont {van Kerkwijk}, \citenamefont {Brett}, \citenamefont {Haldane}, \citenamefont {del R{\'{i}}o}, \citenamefont {Wiebe}, \citenamefont {Peterson}, \citenamefont {G{\'{e}}rard-Marchant}, \citenamefont {Sheppard}, \citenamefont {Reddy}, \citenamefont {Weckesser}, \citenamefont {Abbasi}, \citenamefont {Gohlke},\ and\ \citenamefont {Oliphant}}]{harris2020array}%
  \BibitemOpen
  \bibfield  {author} {\bibinfo {author} {\bibfnamefont {C.~R.}\ \bibnamefont {Harris}}, \bibinfo {author} {\bibfnamefont {K.~J.}\ \bibnamefont {Millman}}, \bibinfo {author} {\bibfnamefont {S.~J.}\ \bibnamefont {van~der Walt}}, \bibinfo {author} {\bibfnamefont {R.}~\bibnamefont {Gommers}}, \bibinfo {author} {\bibfnamefont {P.}~\bibnamefont {Virtanen}}, \bibinfo {author} {\bibfnamefont {D.}~\bibnamefont {Cournapeau}}, \bibinfo {author} {\bibfnamefont {E.}~\bibnamefont {Wieser}}, \bibinfo {author} {\bibfnamefont {J.}~\bibnamefont {Taylor}}, \bibinfo {author} {\bibfnamefont {S.}~\bibnamefont {Berg}}, \bibinfo {author} {\bibfnamefont {N.~J.}\ \bibnamefont {Smith}}, \bibinfo {author} {\bibfnamefont {R.}~\bibnamefont {Kern}}, \bibinfo {author} {\bibfnamefont {M.}~\bibnamefont {Picus}}, \bibinfo {author} {\bibfnamefont {S.}~\bibnamefont {Hoyer}}, \bibinfo {author} {\bibfnamefont {M.~H.}\ \bibnamefont {van Kerkwijk}}, \bibinfo {author} {\bibfnamefont {M.}~\bibnamefont {Brett}}, \bibinfo {author} {\bibfnamefont
  {A.}~\bibnamefont {Haldane}}, \bibinfo {author} {\bibfnamefont {J.~F.}\ \bibnamefont {del R{\'{i}}o}}, \bibinfo {author} {\bibfnamefont {M.}~\bibnamefont {Wiebe}}, \bibinfo {author} {\bibfnamefont {P.}~\bibnamefont {Peterson}}, \bibinfo {author} {\bibfnamefont {P.}~\bibnamefont {G{\'{e}}rard-Marchant}}, \bibinfo {author} {\bibfnamefont {K.}~\bibnamefont {Sheppard}}, \bibinfo {author} {\bibfnamefont {T.}~\bibnamefont {Reddy}}, \bibinfo {author} {\bibfnamefont {W.}~\bibnamefont {Weckesser}}, \bibinfo {author} {\bibfnamefont {H.}~\bibnamefont {Abbasi}}, \bibinfo {author} {\bibfnamefont {C.}~\bibnamefont {Gohlke}},\ and\ \bibinfo {author} {\bibfnamefont {T.~E.}\ \bibnamefont {Oliphant}},\ }\bibfield  {title} {\emph {\bibinfo {title} {Array programming with {NumPy}}},\ }\href {https://doi.org/10.1038/s41586-020-2649-2} {\bibfield  {journal} {\bibinfo  {journal} {Nature}\ }\textbf {\bibinfo {volume} {585}},\ \bibinfo {pages} {357--362} (\bibinfo {year} {2020})}\BibitemShut {NoStop}%
\bibitem [{\citenamefont {Shannon}(1948)}]{shannon1948mathematical}%
  \BibitemOpen
  \bibfield  {author} {\bibinfo {author} {\bibfnamefont {C.~E.}\ \bibnamefont {Shannon}},\ }\bibfield  {title} {\emph {\bibinfo {title} {A mathematical theory of communication}},\ }\href@noop {} {\bibfield  {journal} {\bibinfo  {journal} {The Bell system technical journal}\ }\textbf {\bibinfo {volume} {27}},\ \bibinfo {pages} {379--423} (\bibinfo {year} {1948})}\BibitemShut {NoStop}%
\bibitem [{\citenamefont {Pelofske}(2025{\natexlab{b}})}]{pelofske_2025_17316067}%
  \BibitemOpen
  \bibfield  {author} {\bibinfo {author} {\bibfnamefont {E.}~\bibnamefont {Pelofske}},\ }\href {https://doi.org/10.5281/zenodo.17316067} {10.5281/zenodo.17316067} (\bibinfo {year} {2025}{\natexlab{b}})\BibitemShut {NoStop}%
\end{thebibliography}%

\end{document}